\documentclass[traditabstract]{aa} 
\usepackage{graphicx}
\usepackage{txfonts}
\usepackage{bm}
\usepackage{color}
\usepackage{subfigure}
\usepackage{multirow}
\usepackage{natbib}
\bibpunct{(}{)}{;}{a}{}{,} 
\newcommand{\sigmav}{\ensuremath{\langle \sigma v \rangle}}
\newcommand{\clumpi}{\mbox{1FGL J0030.7+0724}}
\def\mathbi#1{\textbf{\em #1}}
\defcitealias{2011APh....34..608H}{H.E.S.S. Collaboration 2011}
\begin{document}
   \title{Dark matter subhaloes as gamma-ray sources \\ and candidates in the first \textit{Fermi}-LAT catalogue}

   \author{H.-S. Zechlin\inst{\ref{inst1}}\thanks{Send offprint requests to: Hannes-S.~Zechlin, \email{hzechlin@physik.uni-hamburg.de}}
	  \and
          M.~V. Fernandes\inst{\ref{inst1}}
	  \and
          D. Els\"asser\inst{\ref{inst2}}
	  \and
	  D. Horns\inst{\ref{inst1}}
}

   \institute{University of Hamburg, Institut f\"ur
     Experimentalphysik, Luruper Chaussee 149, D-22761 Hamburg,
     Germany\label{inst1} \and University of W\"urzburg, Institut
     f\"ur Theoretische Physik und Astrophysik, Am Hubland, D-97074
     W\"urzburg, Germany\label{inst2} }

\date{Received 7 July 2011 / Accepted 3 November 2011}

\abstract{The standard paradigm of hierarchical structure formation in
  a $\Lambda$CDM universe predicts the presence of dark matter
  subhaloes, hosted by Milky Way-sized galaxies. Anticipated subhalo
  masses range from $10^{10}$ down to a cut-off mass between $10^{-3}$
  and $10^{-11}\,\mathrm{M}_\odot$. If dark matter is composed of
  heavy self-annihilating or decaying particles, these subhaloes could
  be visible in the $\gamma$-ray band as faint and temporally constant
  sources without astrophysical counterparts. Based upon realistic
  subhalo models and current observational constraints on annihilating
  dark matter scenarios, we predict that one massive Galactic subhalo
  between $10^6$ and $10^8\,\mathrm{M}_\odot$ may already be present
  in the 11-month catalogue of \textit{Fermi}-LAT. Indeed, at least
  twelve objects in the first \textit{Fermi} catalogue qualify as
  candidates. The most promising object, \object{\clumpi}, is
  investigated in detail using a dedicated \textit{Swift} X-ray
  follow-up observation and a refined positional analysis of the
  24-month \textit{Fermi}-LAT data. With the new observations, seven
  point-like X-ray sources have been discovered, of which
  \object{SWIFT~J003119.8+072454}, which coincides with a faint radio
  source (12\,mJy at 1.4\,GHz), serves as a counterpart candidate of
  \clumpi. The broad-band spectral energy distribution is consistent
  with a high-energy-peaked blazar. However, flux and extent of
  \clumpi\space may also be compatible with a dark matter
  subhalo. Detection of temporal variability or improved astrometry of
  \clumpi\space are necessary to rule out or confirm an astrophysical
  origin. We discuss strategies to identify $\gamma$-ray sources that
  are associated with self-annihilating dark matter subhaloes.}


   \maketitle
%

\section{Introduction}
Several astrophysical observations indicate that in the early as well
as in the present Universe a non-baryonic form of dark matter (DM)
prevails over the baryonic matter content. Structure formation favours
a cold dark matter (CDM) scenario \citep[for recent reviews see,
  e.g.,][]{2005PhR...405..279B,2009arXiv0907.1912D,2010Natur.468..389B}. However,
the nature of DM remains unknown. A class of promising candidates for
CDM are stable, weakly interacting, massive particles (WIMPs) with
masses between 10 and $10^{5}$\,GeV, predicted by theories that extend
the standard model of particle physics. The most prominent extensions
encompass those based on supersymmetry and universal extradimensions,
which were invented to solve inconsistencies of the standard model at
high energy scales [$\mathcal{O}(\mathrm{TeV})$], and which deliver
adequate DM candidates in this way. These particles can
self-annihilate or decay, producing detectable signatures in the final
states such as energetic photons ($\gamma$ rays), antimatter, and
leptons.

Unravelling the nature of DM remains a challenging problem for
astronomy and particle physics, and a variety of attempts to detect
signals have been made, using both direct and indirect detection
techniques. For instance, multi-wavelength observations of
astrophysical targets have constrained the self-annihilation rate of
DM, which is related to the thermally averaged annihilation cross
section. In particular, regions with high DM densities such as the
Galactic Centre
\citep{2006PhRvL..97v1102A,2006PhRvL..97x9901A,2010arXiv1012.2292M,2011PhRvL.106p1301A},
Galactic Ridge \citep{2006Natur.439..695A}, dwarf spheroidal galaxies
\citep[dSphs;][\citetalias{2011APh....34..608H}]{2007PhRvD..75b3513C,2008ApJ...679..428A,2008APh....29...55A,2009ApJ...691..175A,2009ApJ...697.1299A,2010APh....33..274A,2010ApJ...712..147A,2010ApJ...720.1174A,2011JCAP...06..035A},
as well as globular \citep{2006A&A...455...21C,2009arXiv0910.4563W,2011ApJ...735...12A}
and galaxy clusters
\citep{2010ApJ...710..634A,2010JCAP...05..025A}
serve as excellent targets. Furthermore, DM annihilation in the entire
Galactic halo as well as its subhalo population produces a diffuse
$\gamma$-ray flux, which contributes to the overall diffuse signal of
the Galaxy. Comparatively stringent upper limits on the annihilation
cross section have been obtained from the combination of both
dedicated observations and bounds obtained from the diffuse
$\gamma$-ray flux
\citep{2010JCAP...11..041A,2010NuPhB.840..284C,2010NuPhB.831..178M,2010JCAP...03..014P,2010arXiv1012.0588Z}.

Based on the theory of hierarchical structure formation, DM haloes of
Milky Way-sized galaxies are anticipated to host numerous DM subhaloes
with masses between a cut-off scale $10^{-11}\!-\!10^{-3}$ and
$10^{10}\,\mathrm{M}_\odot$ \citep[e.g.,][]{2009NJPh...11j5027B},
where $\mathrm{M}_\odot$ denotes the solar mass. This expectation is a
consequence of the early collapse of overdensities in the expanding
Universe \citep{2005Natur.433..389D}, leading to the formation of
initially low-mass haloes, which subsequently serve as building-blocks
for larger haloes by merging at later times. Besides analytical
calculations
\citep[e.g.,][]{2003PhRvD..68j3003B,2006PhRvD..73f3504B,2008PhRvD..77h3519B},
recent numerical high-resolution $N$-body simulations of structure
formation in a $\Lambda$CDM cosmology \citep{2011ApJS..192...18K},
such as the Aquarius Project
\citep{2008MNRAS.391.1685S,2008Natur.456...73S} or the Via Lactea II
simulation \citep{2008Natur.454..735D,2009MNRAS.394..641Z}, allow to
study substructures in detail. For a Milky Way-type galaxy, these
simulations predict a large number of subhaloes (up to $10^{16}$) with
masses distributed following a power law, $\mathrm{d}N/\mathrm{d}M
\propto M^{-\alpha}$, where $\alpha \in [1.9;2.0]$. The DM density
profiles of large-mass subhaloes are found to be similar to the
host's, which results in high central densities
\citep{2008MNRAS.391.1685S}. The spatial distribution of subhaloes is
``anti-biased'', i.e., the dominant fraction is placed far away from
the host-halo's centre.

In self-annihilating DM scenarios, subhaloes are expected to appear as
weak point-like or moderately extended $\gamma$-ray sources, and a
small fraction of them could be detectable with current high- or very
high-energy (VHE) $\gamma$-ray telescopes
\citep[e.g.,][]{2005PhRvL..95u1301P,2008MNRAS.384.1627P,2011PhRvD..83b3518P,2008Natur.456...73S,2008ApJ...686..262K,2009PhRvD..80b3520A,2010PhRvD..82f3501B,2011PhRvD..83a5003B,2011arXiv1110.6868Z}. Examples
for currently operating telescopes are the \textit{Fermi}-LAT
\citep[$20\,\mathrm{MeV} - 300\,\mathrm{GeV}$,][]{2009ApJ...697.1071A}
and imaging air Cherenkov telescopes (IACTs; $E\gtrsim 100$\,GeV) such
as H.E.S.S. \citep{2006A&A...457..899A}, MAGIC
\citep{2008ApJ...674.1037A,2010NIMPA.623..437F}, and VERITAS
\citep{2002APh....17..221W}. In the near future, a significant
improvement in the overall sensitivity and lower energy threshold will
be achieved by upcoming experiments such as H.E.S.S.-II
\citep{2005ICRC....5..163V} and CTA
\citep{2010NIMPA.623..408F,2010arXiv1008.3703C,2011NIMPA.630..285D}. Such
instruments are possibly sufficiently sensitive to detect nearby
large-mass subhaloes of $\mathcal{O}(10^6)\,\mathrm{M}_\odot$ within
distances of $\mathcal{O}(1)\,\mathrm{kpc}$.

This paper consists of two separate parts, which can in principle be
read independently. The first part (Sects. \ref{sect:gamma_rays} to
\ref{sect:Fermi_clumps}) investigates the detectability of subhaloes
with \textit{Fermi}-LAT, where the basic framework for predicting the
$\gamma$-ray properties of subhaloes is laid out in Sects.
\ref{sect:gamma_rays} and \ref{sect:candidate_sources}. In Sect.
\ref{sect:Fermi_clumps}, properties of detectable subhaloes are
investigated by means of a fiducial source. In the second part,
Sect. \ref{sect:1FGL_searches} discusses the search for DM subhaloes
in the first \textit{Fermi}-LAT point-source catalogue (1FGL) and
subsequent multi-wavelength studies of the most promising candidate,
\clumpi. A discussion of the physical origin of \clumpi\space and
prospects for IACTs are presented in Sect. \ref{sect:discussion}.

Throughout this paper, Hubble's constant is $H_0 =
73\,\mathrm{km}\,\mathrm{s}^{-1}\,\mathrm{Mpc}^{-1}$, yielding the
present value of the Universe's critical density $\rho_\mathrm{crit} =
3H_0^2/(8 \pi G_\mathrm{N}) \simeq 1.48 \times
10^{11}\,\mathrm{M}_\odot\,\mathrm{Mpc}^{-3}$, where $G_\mathrm{N}$
denotes Newton's gravitational constant
\citep[e.g.,][]{2007ApJS..170..377S}.

\section{Gamma rays from DM subhaloes} \label{sect:gamma_rays}
In the following, the $\gamma$-ray flux from DM subhaloes will be
derived, based upon current theoretical models of the corresponding
radial density distribution.

With respect to undisturbed, isolated galactic haloes, henceforth
field haloes, the general formation history of (embedded) subhaloes
differs significantly. Analytical models and numerical $N$-body
simulations of structure formation found their physical properties to
depend on particular evolutionary conditions, i.e., formation time,
evolution, and orbit \citep[see][and references
  therein]{2007ApJ...667..859D,2008ApJ...679.1680D}. Tidal interaction
with the gravitational potential of the host halo leads to tidal
stripping and heating, and can therefore truncate the outer region of
subhaloes. In the following, two different approaches will be
discussed. On the one hand, subhaloes are modelled assuming negligible
tidal effects and are therefore considered to be in a genuine
virialised state. Because this approximation is (at least) valid for
field haloes, this model will be tagged as \textit{field-halo model}
(FHM). On the other hand, a second and more realistic model is
considered to account for subhalo evolution, henceforth referred to as
\textit{subhalo model} (SHM).

\subsection{Density profile} \label{sect:density}
The subhalo's DM density profile $\rho(r)$ is assumed to follow
\begin{equation} \label{eq:sub_profile}
 \rho(r) = \frac{\rho_\mathrm{s}}{(r/r_\mathrm{s})^\gamma \left( 1+
   r/r_\mathrm{s} \right)^2} \left\{ \begin{array}{lcl} 1 &
   \mathrm{for} & r \leq r_\mathrm{cut}, \\ 
   0 & \mathrm{for} & r > r_\mathrm{cut}, \end{array} \right.
\end{equation}
where $r$ denotes the distance to the subhalo's centre. In general,
the profile cuts at an outer radius $r_\mathrm{cut}$, which is the
virial or tidal radius ($R_\mathrm{vir}$ or $R_\mathrm{t}$),
respectively. Given $\gamma = 1.0$ for the remainder, the profile
follows the universal spherically symmetric Navarro-Frenk-White (NFW)
profile, well-fitting haloes resolved in numerical
simulations\footnote{Note that details on the very inner slope of halo
  profiles remain to be clarified, by simulations as well as
  observationally
  \cite[e.g.,][]{2011ApJ...733L..46W,2007MNRAS.378...41S}.}
\citep{1997ApJ...490..493N}. The profile is defined by two parameters:
a characteristic inner radius $r_\mathrm{s}$, where the effective
logarithmic slope of the profile is $-2$, and an inner density $\rho_s
= 4\rho(r_\mathrm{s})$. In case of FHM haloes, which are not subject
to tides, both parameters are related to each other by the virial halo
mass $M_\mathrm{vir}$. This quantity is defined as the mass inside the
sphere of radius $R_{\mathrm{vir}}$, which encloses a mean density of
$\Delta_{\mathrm{c}}$ times the critical density of the Universe at
the considered redshift $z$
\citep{1997ApJ...490..493N,2001MNRAS.321..559B}, $M_\mathrm{vir} :=
4\pi/3 \,\Delta_\mathrm{c} \rho_\mathrm{crit} R^3_\mathrm{vir}$. The
virial overdensity at $z = 0$ is $\Delta_\mathrm{c} \approx 100$, as
suggested by models of the dissipationless spherical top-hat collapse
\citep{1996MNRAS.282..263E,1998ApJ...495...80B} and assuming present
concordance cosmology. In general, the subhalo mass $M$ is given by a
volume integration of Eq. \ref{eq:sub_profile}, revealing $M = 4\pi
\rho_\mathrm{s} r^3_\mathrm{s} f(c)$, where $f(c) \equiv
\ln(1+c)-c/(1+c)$ and $c$ denotes the concentration parameter of the
subhalo. For non-disturbed haloes, the concentration is then given by
the virial concentration $c_\mathrm{vir} \equiv
R_\mathrm{vir}/r_\mathrm{s}$. Generally, the concentration depends on
the subhalo mass and redshift, $c= c(M,z)$, where lighter haloes have
higher concentrations
\citep{1996ApJ...462..563N,1997ApJ...490..493N,2001MNRAS.321..559B}. Since
observational estimates are lacking \citep[see Sect. 2.2.1 in][and
  references therein]{2008A&A...479..427L}, $c(M)$ is adopted from
$N$-body simulations. For the FHM, the toy model of
\citet{2001MNRAS.321..559B}\footnote{This model extends a proposal by
  \citet{1997ApJ...490..493N}.} is used, where the halo's (average)
virial concentration at redshift $z$ is connected with the density of
the Universe at the halo's (mass-dependent) collapse redshift
$z_c(M)$, $c_\mathrm{vir}=K (1+z_c)/(1+z)$. The contraction parameter
$K$ is constant and independent of cosmology. To determine the mass
dependence of $z_{c}$ and, therefore, of $c_\mathrm{vir}$ at $z=0$,
the low-mass extrapolation of the Bullock model by
\citet{2008A&A...479..427L} is adapted. Because $c_\mathrm{vir}$
implicitly depends on $\Delta_\mathrm{c}$ (see also
Sect. \ref{sect:luminosity}), a conversion of $c_\mathrm{vir}$ to
$\Delta_\mathrm{c}=100$ was
applied\footnote{\citeauthor{2008A&A...479..427L} chose
  $\Delta_\mathrm{c}=81.6$.} with the relation of
\citet{2003ApJ...584..702H}. The concentration-to-mass relation is
well-fitted by the polynomial form
\begin{equation} \label{eq:cvir_lavalle}
 \ln (c^\mathrm{FHM}_\mathrm{vir}) = \sum_{i=0}^{4} c^\mathrm{FHM}_i
 \times \left[ \ln \left( \frac{M}{\mathrm{M}_\odot} \right)
   \right]^i,
\end{equation}
$c^\mathrm{FHM}_i =
\{4.265,-0.0384,-3.91\times10^{-4},-2.2\times10^{-6},-5.5\times10^{-7}\}$.
Note that this model almost equals the relation derived by
\citet{2011PhRvD..83b3518P} for a cosmology as used in the Aquarius
simulation. Regarding the concentration of SHM haloes, the low-mass
extrapolation of the Bullock model provides a conservative estimate
\citep[cf.,][]{2008MNRAS.384.1627P}.

\begin{figure}
  \resizebox{\hsize}{!}{\includegraphics{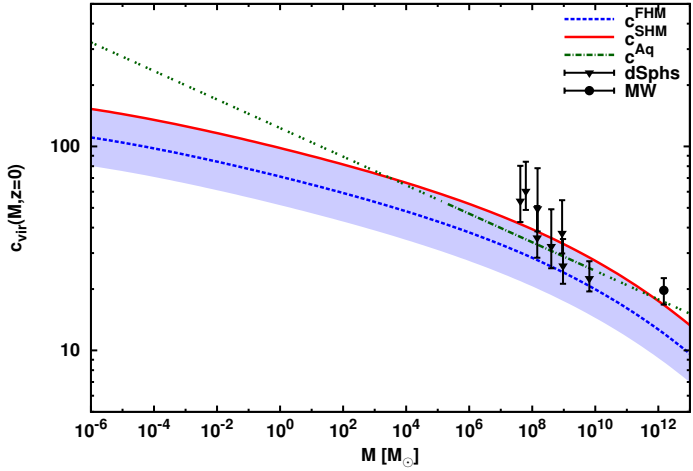}}
  \caption{Concentration-to-mass relation for subhaloes at $z=0$. The
    dashed blue line depicts the concentration predicted by the
    Bullock model (FHM), while its corresponding scatter is given by
    the blue-shaded area. The mean concentration corrected for subhalo
    evolution (SHM) is shown by the solid red line for a
    galactocentric distance of $74\,\mathrm{kpc}$, the average
    distance of the dSph galaxies included from
    \citet{2010ApJ...712..147A} (black triangles). For comparison, the
    concentration derived for Aquarius subhaloes is indicated by the
    dot-dashed dark green line within its validity range, the
    low/high-mass extrapolation by the double-dotted green line. The
    black filled circle marks the virial concentration of the MW.}
  \label{fig:conc_v}
\end{figure}

However, because subhalo formation differs from that of field haloes
and includes tidal truncation at $R_\mathrm{t}$, the virial
concentration is not well defined for subhaloes\footnote{In general,
  the physical subhalo radius $R_\mathrm{t}$ is smaller than the
  formally defined virial radius $R_\mathrm{vir}$, implying the
  physical subhalo mass $M_\mathrm{t}$ to be smaller than
  $M_\mathrm{vir}$. For massive subhaloes, the
  $M_\mathrm{t}(M_\mathrm{vir})$ relation is approximately linear,
  where $M_\mathrm{t}/M_\mathrm{vir} \approx 0.2$ (see Appendix
  \ref{app:cvir_Aq}). Given the empirical model correction discussed
  below, the formal virial quantities will be used in the remainder of
  the paper.}  \citep{2007ApJ...667..859D}. Therefore, the SHM
incorporates an empirical correction of $c_\mathrm{vir}$. Indicated by
numerical simulations, the concentration of subhaloes increases with
decreasing distance to the host's centre $D_\mathrm{gc}$,
\begin{equation} \label{eq:cSHM}
 c^\mathrm{SHM}_\mathrm{vir}(M,D_\mathrm{gc}) =
 c^\mathrm{FHM}_\mathrm{vir}(M) \left(
 \frac{D_\mathrm{gc}}{R^\mathrm{MW}_\mathrm{vir}} \right)^{-\alpha_D},
\end{equation}
see \citet{2007ApJ...667..859D,2008ApJ...679.1680D} and
\citet{2008ApJ...686..262K}. The galactocentric distance is
\mbox{$D_\mathrm{gc} = ( R_0^2 + D^2 - 2 R_0 D \cos l \cos b
  )^{1/2}$}, where $D$ denotes the subhalo's distance to the Sun,
$(l,b)$ its position in galactic coordinates, and $R_0=(8.28 \pm
0.29)\,\mathrm{kpc}$ the Sun's distance to the Galactic Centre
\citep{2010JCAP...08..004C}. The virial radius of the Milky Way (MW)
is $R^\mathrm{MW}_\mathrm{vir} =
c^\mathrm{MW}_\mathrm{vir}\,r_\mathrm{s}^\mathrm{MW} \approx (288 \pm
61)\,\mathrm{kpc}$, where $c^\mathrm{MW}_\mathrm{vir} = 19.70 \pm
2.92$ and $r^\mathrm{MW}_\mathrm{s} = (14.65 \pm 2.24)\,\mathrm{kpc}$
\citep{2010JCAP...08..004C}. The power-law slope $\alpha_D = 0.237$ is
adopted as fitting subhaloes resolved in the Aquarius simulation
\citep{2011PhRvD..83b3518P}.

Intrinsic to the stochastic process of halo formation, the
concentration of individual haloes scatters around the median
$\overline{c}$ provided by the quantities
$c^\mathrm{FHM}_\mathrm{vir}(M)$ and
$c^\mathrm{SHM}_\mathrm{vir}(M,D_\mathrm{gc})$, respectively. The
corresponding probability distribution follows a lognormal,
\begin{equation} \label{eq:scatter}
 P(c,\overline{c}) = \frac{\log_{10} e}{\sqrt{2 \pi} \sigma_{\log_{10}
     c} \,c} \exp \left[ -\frac{1}{2} \left( \frac{\log_{10} c -
     \log_{10} \overline{c}}{\sigma_{\log_{10} c}} \right)^2 \right],
\end{equation} 
where $\sigma_{\log_{10} c} = 0.14$ \citep{2001MNRAS.321..559B,2002ApJ...568...52W}.
 
The concentration-to-mass relations are shown in
Fig. \ref{fig:conc_v}. For the FHM, the scatter is also depicted
(68\%\,c.l. of $\log_{10} c$), see Eq. \ref{eq:scatter}. In addition
to the concentration of the MW, Fig. \ref{fig:conc_v} contains a
selection of eight dSphs that are associated with sufficiently precise
stellar data, which allow a conclusive modelling of the DM
distribution \citep[see][]{2010ApJ...712..147A}. Each dSph is modelled
with a NFW profile with parameters chosen to fit measurements of
stellar line-of-sight velocities and their distributions \citep[see
  also][]{2009JCAP...06..014M}. The dSph's virial concentration is
given by its characteristic density,
$\rho_\mathrm{s}=\Delta_\mathrm{c} \rho_\mathrm{crit}
c_\mathrm{vir}^3/[3f(c_\mathrm{vir})]$, where tidal effects on the
inner system are assumed to be negligible. The SHM is depicted for
$D_\mathrm{gc} = 74\,\mathrm{kpc}$, the average galactocentric
distance of the dSph subset. Additionally, the models are confronted
with direct predictions of the Aquarius simulation, derived from
scaling relations fitting subhaloes observed in the
simulation. Details are provided in Appendix \ref{app:cvir_Aq}.

Within its scatter, the concentration model of FHM haloes consistently
describes the DM profile of dSph galaxies and the MW itself. However,
the median values $c^\mathrm{FHM}_\mathrm{vir}$ underpredict dSphs,
whereas the subhalo model SHM provides convincing agreement (as
expected by $N$-body simulations). The concentration derived directly
from the Aquarius simulation confirms the SHM within the validity
range, see Fig. \ref{fig:conc_v}. Note that the mean distance of
subhaloes resolved in Aquarius is 64\,kpc.

\subsection{DM annihilation in subhaloes} \label{sect:luminosity}
For self-annihilating particles, the total rate of photons (or
particles) emitted by a DM subhalo with energy $E$ in the interval
$[E_1;E_2]$ is
\begin{equation} \label{eq:luminosity}
 \mathcal{L} = \frac{\sigmav_\mathrm{eff} N_\gamma}{2 m_\chi^2} \int
 \mathrm{d}V \rho^2(r) \propto \frac{M^2}{r_\mathrm{s}^3 f(c)^2},
 \quad N_\gamma = \hskip -1mm \int\limits_{E_1/m_\chi}^{E_2/m_\chi}
 \mathrm{d}x \frac{\mathrm{d}N_\gamma}{\mathrm{d}x},
\end{equation}
where $\sigmav_\mathrm{eff}$ is the thermally averaged annihilation
cross section times the relative velocity, $m_\chi$ the WIMP mass, and
$\mathrm{d}N_\gamma/\mathrm{d}x$, $x \equiv E/m_\chi$, denotes the
differential spectrum of photons per annihilation. Assuming
$r_\mathrm{s} \ll D$, the produced photon flux is given by
$\phi=\mathcal{L}/(4\pi D^2)$. The solution of the integral holds for
$\gamma = 1.0$ and $c \gg 1$. In Eq. \ref{eq:luminosity}, a small,
flat core replacing the unphysical singularity at the halo centre
\citep{1992PhLB..294..221B} is safely neglected (given the NFW profile
used here). For $\gamma = 1.0$, Eq. \ref{eq:luminosity} simplifies via
\mbox{$r_\mathrm{s} = [ 3 M/(4\pi \Delta_\mathrm{c} \rho_\mathrm{crit}
    c_\mathrm{vir}^3) ]^{1/3}$}:
\begin{equation} \label{eq:luminosity_vir}
 \mathcal{L} = \frac{\sigmav_\mathrm{eff} N_\gamma \Delta_\mathrm{c}
   \rho_\mathrm{crit}}{18 m_\chi^2} \frac{ M
   c_\mathrm{vir}^3}{f(c_\mathrm{vir})^2}.
\end{equation}
For a $\gamma=1.2$ profile (Eq. \ref{eq:sub_profile}) the
photon rate increases by a factor of \mbox{$\sim 1.5$} for subhaloes
above $10^3\,\mathrm{M}_\odot$. DM annihilation in subhaloes may be
additionally boosted by sub-substructure populations
\citep[see][]{2007PhRvD..75h3526S,2008ApJ...686..262K,2009JCAP...06..014M}. Conveniently,
the value of $\sigmav_\mathrm{eff}$ is normalised to the value
$\sigmav_0 = 3 \times 10^{-26}\,\mathrm{cm}^3 \mathrm{s}^{-1}$, which
leads to the correct relic density. An increase of the annihilation
rate, a so-called boost factor $\sigmav_\mathrm{eff}/\sigmav_0$, could
in principle be related to the underlying particle physics framework
\citep[e.g,][]{2004PhRvD..70j3529F} and effects such as Sommerfeld
enhancement \citep[e.g.,][]{2009PhRvD..79a5014A,2009Sci...325..970K}.

\begin{figure}[t]
  \resizebox{\hsize}{!}{\includegraphics{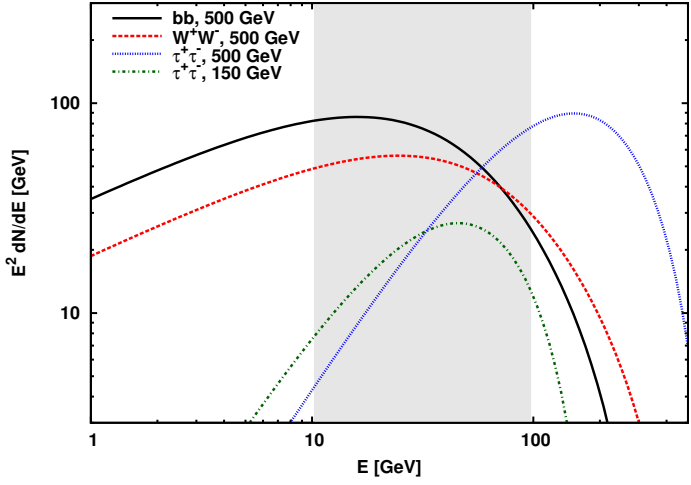}}
  \caption{Differential $\gamma$-ray energy spectra $E^2
    \frac{\mathrm{d}N}{\mathrm{d}E}$ originating from final-state
    fragmentation of WIMP annihilation. Four different WIMP models are
    depicted: mass \mbox{$m_\chi = 500\,\mathrm{GeV}$}, final states:
    $b\overline{b}$ (solid black line), $W^+W^-$ (dashed red line),
    $\tau^+\tau^-$ (dotted blue line); mass \mbox{$m_\chi =
      150\,\mathrm{GeV}$}, final state $\tau^+\tau^-$ (dot-dashed
    green line). The parametrisations are valid down to $E/m_\chi
    \approx 0.01$. The grey-shaded area indicates the considered
    energy range $E \in [10;100]\,\mathrm{GeV}$.  }
  \label{fig:DM_spectra}
\end{figure}

We considered DM to be composed of self-annihilating heavy WIMPs of
mass $m_\chi = 500\,\mathrm{GeV}$ and investigated three distinct
annihilation models: two of them with total annihilation in heavy
quarks or gauge bosons ($b\overline{b}$ and $W^+W^-$) and a model with
total annihilation in the leptons $\tau^+\tau^-$. Additionally, we
considered WIMPs of $m_\chi = 150\,\mathrm{GeV}$ for annihilation in
$\tau^+\tau^-$ final states. Particles of this type are, for instance,
provided by supersymmetric theories, e.g., manifested in the
neutralino. The WIMP masses chosen are compatible with WIMPs which
might explain the recently observed cosmic-ray electron and positron
excess, see, e.g., \citet{2010NuPhB.831..178M}. In general,
\textit{heavy} WIMPs are also supported by collider searches such as
the non-detection of supersymmetric particles in the 7\,TeV run of the
Large Hadron Collider (for an integrated luminosity of 35\,pb$^{-1}$)
\citep[][and references
  therein]{2011EPJC...71.1682A,2011arXiv1105.3152C}.

Given these final annihilation states, hadronisation and the
subsequent decay of $\pi^0$-mesons lead to a continuous $\gamma$-ray
spectrum. The resulting photon spectra
$\mathrm{d}N_\gamma/\mathrm{d}x$ were modelled using parametrisations
provided by \citet{2004PhRvD..70j3529F}, see
Fig. \ref{fig:DM_spectra}. Note that photons produced by final state
radiation (FSR) and virtual internal bremsstrahlung (VIB)
\citep{2005PhRvL..95x1301B,2005PhRvL..94m1301B,2008JHEP...01..049B}
are neglected, because a significant contribution of FSR is only
expected for $W^+W^-$ at high energies ($E > 0.6\,m_\chi$) and
possible contributions of VIB are highly model-dependent.

\section{Candidate gamma-ray sources} \label{sect:candidate_sources}
Given a high WIMP mass, a DM subhalo will show up as steady
(very)~high-energy $\gamma$-ray source. The differential photon
spectrum follows a hard power law (index $\Gamma \lesssim 1.5$) that
cuts off exponentially at energies beyond $10$\,GeV, see
Fig. \ref{fig:DM_spectra}.

Candidate sources are selected according to their possibility to
originate from DM subhaloes based on their observational quantities
flux and angular extent. Note that the detailed spectral shape of
faint sources is observationally rather unconstrained. Via
Eq. \ref{eq:luminosity_vir}, the effective self-annihilation cross
section $\sigmav_\mathrm{eff}$ required to obtain a given flux $\phi$
for the intrinsic source extent $\theta_\mathrm{s}$ is determined by
$\mathcal{L}=4\pi D^2 \phi$, where $\theta_\mathrm{s}$ constrains the
distance $D$ to the subhalo. For feasible candidate sources the
required $\sigmav_\mathrm{eff}$ should comply with current
observational constraints.

Conveniently, the characteristic profile radius $r_\mathrm{s}$ (see
Eq. \ref{eq:sub_profile}) traces the intrinsic extent of a DM subhalo,
because for an NFW profile $87.5\%$ of the total luminosity
is produced within $r_\mathrm{s}$ (see Table \ref{tab:sigmav_eff} for values of
$r_\mathrm{s}$). Therefore, the distance to the subhalo is $D \approx
r_\mathrm{s}/\theta_\mathrm{s}$, where $\theta_\mathrm{s}$ denotes the
angle corresponding to $r_\mathrm{s}$. Owing to the centrally
peaked profile, about 68\% of the total luminosity is emitted
within the angle $\theta_{68} \simeq 0.46\,\theta_\mathrm{s}$. The
following relations are given with respect to $\theta_\mathrm{s}$ and
can easily be adjusted for $\theta_{68}$, which is more convenient for a
comparison with observational data. With \mbox{$r_\mathrm{s} = [ 3
    M/(4\pi \Delta_\mathrm{c} \rho_\mathrm{crit} c_\mathrm{vir}^3)
  ]^{1/3}$}, the distance to a subhalo with given $\theta_\mathrm{s}$
is related to its mass and concentration. In the FHM,
\begin{equation} \label{eq:D_FHM}
 D_\mathrm{FHM}(M;\theta_\mathrm{s}) \simeq 3.8 \left(
 \frac{M}{10^6\,\mathrm{M}_\odot} \right)^{1/3} \left(
 \frac{c^\mathrm{FHM}_\mathrm{vir}}{37.9} \right)^{-1} \left(
 \frac{\theta_\mathrm{s}}{\mathrm{deg}} \right)^{-1}\,\mathrm{kpc}.
\end{equation}
Note that the concentration depends on the subhalo mass via
Eq.~\ref{eq:cvir_lavalle} as well as, in the SHM, on the object's
position $(l,b)$.

For a given WIMP model $\sigmav_\mathrm{eff}$ is then fully
determined by the subhalo mass (Eq. \ref{eq:luminosity_vir}) and the
observed quantities flux and extent:
\begin{equation} \label{eq:boost}
 \sigmav_\mathrm{eff}(M;\phi,\theta_\mathrm{s}) =
 96\,\pi^{\frac{1}{3}} \frac{m^2_\chi}{N_\gamma} \left( \frac{3}{4
   \Delta_\mathrm{c} \rho_\mathrm{crit}} \right)^{5/3}
 \frac{\phi}{\theta^2_\mathrm{s}} \, \frac{M^{-1/3}
   f(c_\mathrm{vir})^2}{c^5_\mathrm{vir}}.
\end{equation}
Additional contributions to the DM signal from annihilation in the
smooth halo as well as the entire subhalo population were
neglected\footnote{For the fiducial candidate in
  Sect. \ref{sect:Fermi_clumps} this additional contribution is less
  than 1\%.}. The required $\sigmav_\mathrm{eff}$ is highly sensitive
to the (observationally unconstrained) concentration, because
Eq. \ref{eq:boost} roughly depends on $c^{-5}_\mathrm{vir}$.

\section{Interpretation of \textit{Fermi} sources as DM subhaloes} \label{sect:Fermi_clumps}
Based on the study of a fiducial candidate source in
Sect. \ref{sect:fiducial_candidate}, the properties of
\textit{Fermi}-LAT detectable subhaloes are investigated in
Sect. \ref{sect:LATdet_subs}.

\subsection{A fiducial candidate} \label{sect:fiducial_candidate}
\subsubsection{Observational properties} \label{sect:fiducial_candidate:obsprop}
\begin{figure*}[t]
\centering
  \includegraphics[width=90mm]{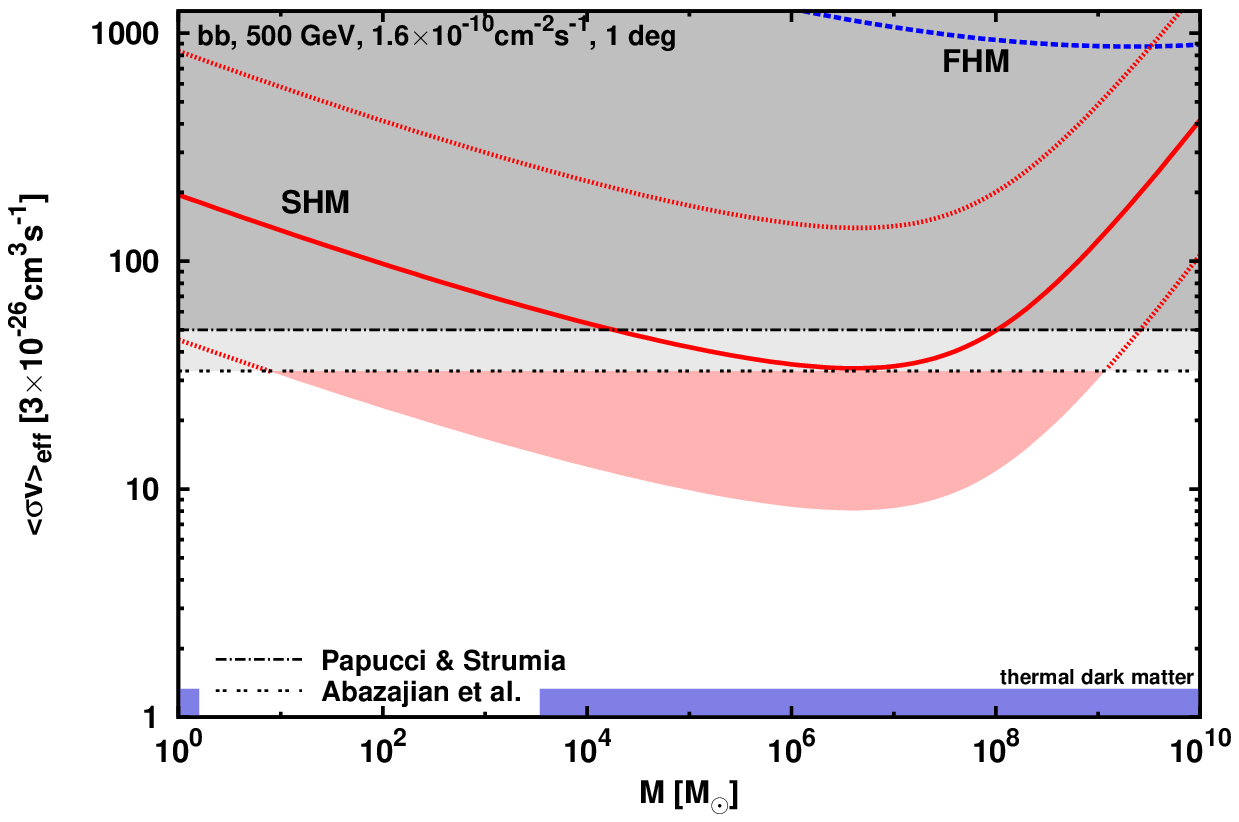}
  \includegraphics[width=90mm]{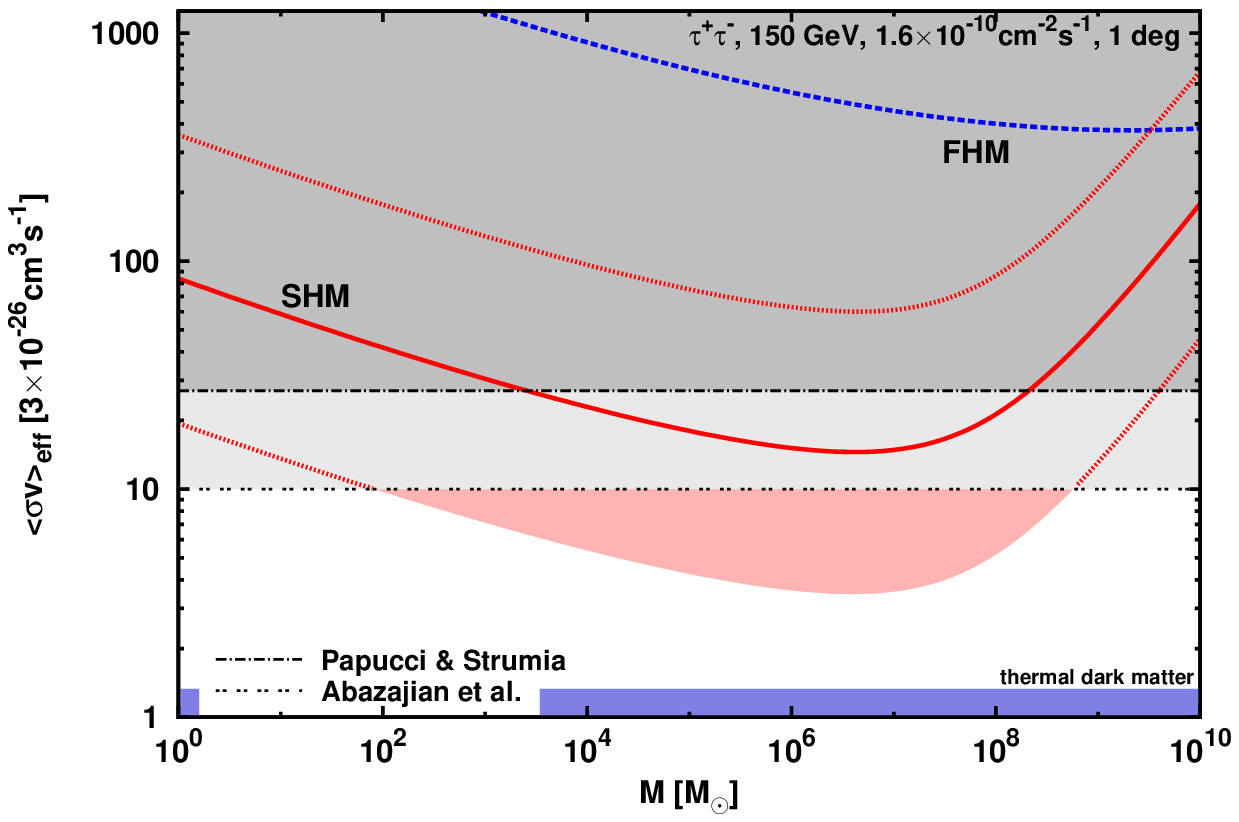}
  \caption{Effective self-annihilation cross section
    $\sigmav_\mathrm{eff}$ required for a moderately extended, faint
    \textit{Fermi}-LAT source to originate from a DM subhalo of mass
    $M$. Assumed source parameters are $\phi(10\!-\!100\,\mathrm{GeV})
    =1.6\times 10^{-10}\,\mathrm{cm}^{-2} \mathrm{s}^{-1}$, an
    intrinsic extent $\theta_\mathrm{s} = 1^\circ$ ($\theta_{68}
    \approx 0.5^\circ$), and the Galactic position
    $(l,b)=(114^\circ,-55^\circ)$. The dashed blue and solid red lines
    indicate the average prediction considering the FHM and SHM,
    respectively. The intrinsic SHM scatter is shown by the red-shaded
    area and the dotted red lines. The \textit{left panel} depicts
    WIMPs of $m_\chi=500\,\mathrm{GeV}$ totally annihilating in
    $b\overline{b}$, while the \textit{right panel} considers
    $m_\chi=150\,\mathrm{GeV}$ with total annihilation in
    $\tau^+\tau^-$. Current contraints on $\sigmav_\mathrm{eff}$ from
    \citet{2010JCAP...03..014P} (grey-shaded) and
    \citet{2010JCAP...11..041A} (light grey-shaded) are plotted in
    combination with the expectation from thermal freeze-out
    (blue-shaded).}
  \label{fig:fiducial}
\end{figure*}
In combination with improving (integrated) sensitivity at high energy
\citep[][and cf., Fig. \ref{fig:spectra}]{2009ApJ...697.1071A}, the
expected energy spectrum of DM subhaloes (Sect. \ref{sect:gamma_rays}
and \ref{sect:candidate_sources}) favours a detection at the
high-energy band of \textit{Fermi}-LAT. In Appendix B, we investigate
the detection sensitivity for faint, moderately extended
($\theta_\mathrm{s} \lesssim 1^\circ$, corresponding to $\theta_{68}
\lesssim 0.5^\circ$), and high-latitude ($|b| > 20^\circ$) sources
between 10 and 100\,GeV in detail. We find that a spectrally hard
high-energy source with a flux $\phi(10\!-\!100\,\mathrm{GeV}) =
1.4\times 10^{-10}\,\mathrm{cm}^{-2}\,\mathrm{s}^{-1}$ and moderate
extent\footnote{In comparison with the point spread function of
  \textit{Fermi}-LAT, \mbox{$\theta_\mathrm{s} = 1^\circ$} (as
  implying \mbox{$\theta_{68} \approx 0.5^\circ$}) corresponds to
  about \mbox{$3\,\sigma_\mathrm{PSF}$}, where
  \mbox{$\sigma_\mathrm{PSF} \approx 0.15^\circ$} for energies beyond
  10\,GeV (see
  http://www-glast.slac.stanford.edu/software/IS/glast\_lat\_performance.htm).}
$\theta_\mathrm{s} = 1^\circ$ can be detected as a point-source with a
reconstructed flux $\phi_\mathrm{p}(10\!-\!100\,\mathrm{GeV}) =
0.9\times 10^{-10}\,\mathrm{cm}^{-2}\,\mathrm{s}^{-1}$, with a
sky-survey exposure of one year. With respect to the true flux $\phi$
emitted by the entire source, the reconstructed flux $\phi_\mathrm{p}$
fitted by the point-source analysis in general decreases with
increasing $\theta_\mathrm{s}$. To account for this effect, the
scaling relation $\phi(\theta_\mathrm{s}) =
h(\theta_\mathrm{s})\,\phi_\mathrm{p}$ is used in Eq. \ref{eq:boost},
where $h(\theta_\mathrm{s}) = 1$ for $\theta_\mathrm{s} \ll
2\,\sigma_\mathrm{PSF}$ and $h(\theta_\mathrm{s}) \approx 0.72 \left(
\theta_\mathrm{s}/\mathrm{deg} \right) + 0.89$ for extended sources up
to \mbox{$\sim 1^\circ$} (see Appendix \ref{app:mc} for details).

The high-energy flux of the fiducial source above 10\,GeV has been
chosen to be at the level of the detection sensitivity,
$\phi^\mathrm{fid}_\mathrm{p}(10\!-\!100\,\mathrm{GeV}) =
10^{-10}\,\mathrm{cm}^{-2}\,\mathrm{s}^{-1}$, assuming an extent of
$\theta^\mathrm{fid}_\mathrm{s} = 1^\circ$. Given the dependence of the SHM
concentration on the galactocentric distance (see Eq.~\ref{eq:cSHM}),
the fiducial source is placed on a particular line-of-sight
chosen to match the location of \clumpi\space (investigated in
Sect. \ref{sect:1FGL_searches}). In general, this line-of-sight serves
as an appropriate (conservative) benchmark, because it points to
(anticentric) positions where the majority of subhaloes is located.

\subsubsection{Subhalo interpretation}
Adopting the properties of the fiducial source,
Fig. \ref{fig:fiducial} depicts the effective enhancement factors
$\sigmav_\mathrm{eff}/\sigmav_0$ required to generate the emission
$\phi^\mathrm{fid}_\mathrm{p}$ by DM annihilation (obtained via
Eq. \ref{eq:boost}). In the left panel, WIMPs of
$m_\chi=500\,\mathrm{GeV}$ are considered to totally annihilate in
$b\overline{b}$, while the right panel assumes
$m_\chi=150\,\mathrm{GeV}$ and annihilation in $\tau^+\tau^-$. For a
given WIMP model, the resulting enhancement factors of the FHM and SHM
are widely different. With respect to the FHM, much less enhancement
is required in the SHM, which is manifested in generically higher
concentrations of SHM subhaloes. Within the scatter of the
concentration intrinsic to the stochastic nature of halo formation
(Eq.~\ref{eq:scatter}), which is shown for the SHM, the necessary
enhancement spans about one order of magnitude. Only moderate
enhancement is required for massive subhaloes between $10^{6}$ and
$10^{7}\,\mathrm{M}_\odot$, where the lowest $\sigmav_\mathrm{eff}$ is
needed for $m_\chi=150\,\mathrm{GeV}$ and $\tau^+\tau^-$ final states
(amongst the WIMP models considered here). The lowest possible
enhancement factors within the concentration scatter of a
$10^6\,\mathrm{M}_\odot$ subhalo are listed in Table
\ref{tab:sigmav_eff} for the different subhalo and the WIMP models of
Sect. \ref{sect:gamma_rays}.

\begin{figure}[t]
  \resizebox{\hsize}{!}{\includegraphics{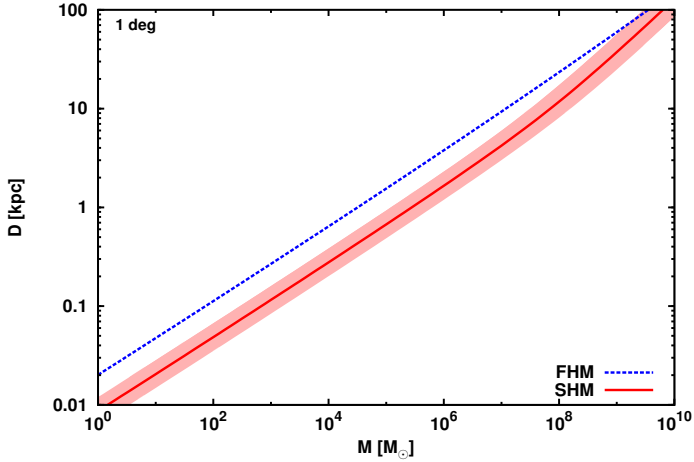}}
  \caption{Distance to the fiducial subhalo in the FHM (dashed blue
    line) and SHM (solid red line), respectively, as function of the
    subhalo mass $M$. The scatter of the SHM distance is indicated by
    the red-shaded area. An extent of $\theta_\mathrm{s} = 1^\circ$
    ($\theta_{68} \approx 0.5^\circ$) and a source position $(l,b) =
    (114^\circ,-55^\circ)$ are assumed.}
  \label{fig:distance}
\end{figure}

The distance to the fiducial candidate anticipated in the FHM and the
SHM is shown in Fig.~\ref{fig:distance}. The intrinsic concentration
scatter implies a corresponding distance scatter for a given halo mass
and angular extent. Note that a similar scatter is present for the
FHM, but is not shown in the figure. Compared with the FHM, tidal
effects lead to higher concentrated subhaloes. This in turn favours a
closer distance at the same mass and angular extent for SHM subhaloes
than for FHM.

\subsubsection{Consistency with observational constraints}
\begin{table}[b]
\caption{\label{tab:sigmav_eff}Enhancement factors
  $\sigmav_\mathrm{eff}/\sigmav_0$ required to explain the fiducial
  $\gamma$-ray source with a DM subhalo of $10^6\,\mathrm{M}_\odot$.}
\centering
\begin{tabular}{lcccccc}
\hline \hline
 \multirow{2}{*}{Model} & \multicolumn{3}{c}{$m_\chi = 500\,\mathrm{GeV}$} &\hspace{-2mm} 150\,GeV & $r^\mathrm{n}_\mathrm{s}$ & \multirow{2}{*}{$g_r(c_\mathrm{vir})$} \\
 & \hspace{-2mm} $b\overline{b}$ & $W^+W^-$ & $\tau^+\tau^-$ & \hspace{-2mm} $\tau^+\tau^-$ & [kpc] & \\
\hline
 FHM &\hspace{-2mm} $\gtrsim 321$ & $\gtrsim 479$ & $\gtrsim 1386$ &\hspace{-2mm} $\gtrsim 138$ & 0.067 & $\left(\frac{c^\mathrm{FHM}_\mathrm{vir}}{37.86}\right)^{-1}$\\
 SHM\tablefootmark{a} & \hspace{-2mm} $\gtrsim 8$ & $\gtrsim 12$ & $\gtrsim 35$ &\hspace{-2mm} $\gtrsim 3$ & 0.029 & $\left(\frac{c^\mathrm{SHM}_\mathrm{vir}}{86.56}\right)^{-1}$\\
\hline
\end{tabular}
\tablefoot{The factors correspond to the WIMP models discussed in the
  text. The respective photon yields are
  $N^{bb\,(WW)\,[\tau\tau]}_\gamma(10\!-\!100\,\mathrm{GeV})=6.95\,(4.66)\,[1.46]$
  for $m_\chi = 500\,\mathrm{GeV}$ and
  $N^{\tau\tau}_\gamma(10\!-\!100\,\mathrm{GeV})= 1.61$ for $m_\chi =
  150\,\mathrm{GeV}$. We list the minimum values within the
  $c$-scatter. In addition, the subhalo's (average) characteristic
  radius $r_\mathrm{s}(M) = r^\mathrm{n}_\mathrm{s} [
    M/(10^6\,\mathrm{M}_\odot)]^{1/3} g_r(c_\mathrm{vir})$ is
  depicted. \tablefoottext{a}{The normalisation of
    $c^\mathrm{SHM}_\mathrm{vir}$ implies the Galactic position
    ($114^\circ$,$-55^\circ$) and intrinsic extent $\theta_\mathrm{s}
    = 1^\circ$, corresponding to $D \approx 1.7$\,kpc ($D_\mathrm{gc}
    \approx 8.8$\,kpc).}  }
\end{table}
The resulting values of $\sigmav_\mathrm{eff}$ can now be checked for
consistency with the diffuse extragalactic $\gamma$-ray background
\citep[EGB;][]{2010PhRvL.104j1101A}. \textit{Fermi}-LAT measurements
of the overall diffuse $\gamma$-ray flux allow the derivation of the
isotropic high-energy EGB, which is shown to be compatible with a
featureless power-law spectrum ($\Gamma= 2.41 \pm 0.05$) and
integrated diffuse flux $\phi_\mathrm{EGB}(>\!100\,\mathrm{MeV}) = (1.03
\pm 0.17) \times 10^{-5}\,\mathrm{cm}^{-2}
\mathrm{s}^{-1}\mathrm{sr}^{-1}$ \citep{2010PhRvL.104j1101A}. The
diffuse $\gamma$-ray flux anticipated from DM annihilation in the
Galactic halo as well as the entire subhalo population is shown in
Fig. \ref{fig:diff} (see Appendix \ref{app:sub_diff} for details), in
comparison with the EGB. Both FHM and SHM subhaloes are depicted for
$m_\chi = 150\,\mathrm{GeV}$ and annihilation in $\tau^+\tau^-$. In
this model, the nearly isotropic diffuse flux from the subhalo
population contributes about 1\% to the EGB (assuming SHM subhaloes
and no sub-substructure) and is fainter than the contribution of the
smooth halo ($\gtrsim\! 3\%$). Note that the flux from the extragalactic
halo population is lower than the contribution of Galactic subhaloes,
see, e.g., \citet{2010JCAP...11..041A}.

The contribution from the smooth halo peaks at the Galactic Centre,
where a high astrophysical foreground is also present, and can
therefore not be isotropic. Given that the EGB has been derived
assuming isotropy, the most robust upper limits on
$\sigmav_\mathrm{eff}$ are determined by the subhalo contribution and
are listed in Table \ref{tab:EGB_ul}, depending on the WIMP model and
cut-off mass. The bounds were obtained requiring that the specific
intensity of the subhalo population $\langle I_\nu(180^\circ,E)
\rangle$ does not exceed the EGB, where $\langle I_\nu(\psi,E)
\rangle$ depends on the angle $\psi$ between the Galactic Centre
direction and line-of-sight and the $\gamma$-ray energy $E$ (see
Appendix \ref{app:sub_diff}).

\begin{table}[t]
\caption{\label{tab:EGB_ul}Upper limits on
  $\sigmav_\mathrm{eff}/\sigmav_0$ from the EGB.}
\centering
\begin{tabular}{lccccc}
\hline \hline
 \multicolumn{2}{c}{WIMP model} & \multicolumn{4}{c}{Upper limit on $\sigmav_\mathrm{eff}/\sigmav_0$} \\
   Channel & $m_\chi$\,[GeV] & FHM & SHM & FHM & SHM \\
\hline
  $b\overline{b}$ & 500 & 1650 & 530 & 1875 & 605\\
  $W^+W^-$ & 500 & 2096 & 673 & 2381 & 769\\
  $\tau^+\tau^-$ & 500 & 3490 & 1121 & 3964 & 1279\\
  $\tau^+\tau^-$ & 150 & 378 & 121 & 429 & 139\\
\hline
 \multicolumn{2}{c}{$M_\mathrm{min}$} & \multicolumn{2}{c}{$10^{-10}\,\mathrm{M}_\odot$} & \multicolumn{2}{c}{$10^{-6}\,\mathrm{M}_\odot$}\\
\hline
\end{tabular}
\tablefoot{We assumed a subhalo mass fraction of
  $f_\mathrm{sh}=15\,\%$ for a cut-off mass $M_\mathrm{min} =
  10^{-6}\,\mathrm{M}_\odot$. Upper limits are listed for the cut-off
  masses bordering a 500\,GeV neutralino scenario
  \citep[see][]{2009NJPh...11j5027B} and with respect to
  $\psi=180^\circ$ and $E=40\,\mathrm{GeV}$. See text and Appendix
  \ref{app:sub_diff} for details.}
\end{table}

However, more stringent constraints have been provided by a more
detailed modelling of the EGB, including all DM components. To
evaluate a possible DM origin of the fiducial source, the results of
\citet{2010JCAP...11..041A} and
\citet{2010JCAP...03..014P}\footnote{In comparison with
  \citeauthor{2010JCAP...03..014P}, the work by
  \citeauthor{2010JCAP...11..041A} includes a fore- and background
  subtraction. Note that the MW halo parameters used by
  \citeauthor{2010JCAP...11..041A} are similar to those adopted in
  this work (see Sect. \ref{sect:density}).} are included in
Fig. \ref{fig:fiducial}. As stated in the introduction, competitive
and similar constraints have been also provided by the non-detection
of various objects with high (central) DM densities. Note, for
instance, that the constraints used here are consistent with recent
bounds from dSph galaxies \citep{2010ApJ...712..147A}.

\begin{figure}[t]
  \resizebox{\hsize}{!}{\includegraphics{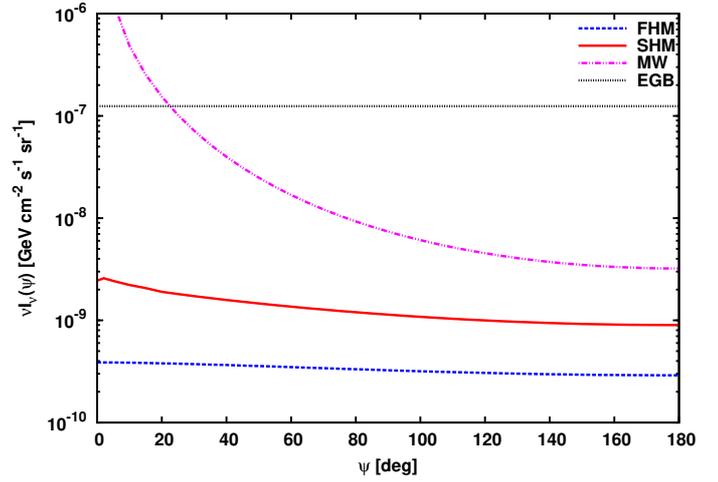}}
  \caption{Average diffuse $\gamma$-ray flux from DM annihilation in
    the Galaxy as function of the angle between the observational and
    Galactic Centre direction $\psi$. The fluxes from the host halo
    (double-dot-dashed magenta line), FHM (dashed blue line), and SHM
    subhaloes (solid red line) are depicted. Each curve was
    derived at the peak energy of $\nu I_\nu(E)$ (40\,GeV) assuming
    total annihilation in $\tau^+\tau^-$ with $m_\chi =
    150\,\mathrm{GeV}$ and $\sigmav_\mathrm{eff} = \sigmav_0$. The
    minimum subhalo mass used was $M_\mathrm{min} =
    10^{-6}\,\mathrm{M}_\odot$ (see Appendix \ref{app:sub_diff}). The
    EGB at 40\,GeV is shown by the dotted black line.}
  \label{fig:diff}
\end{figure}

\subsection{\textit{Fermi}-LAT detectable subhaloes} \label{sect:LATdet_subs}
\subsubsection{Expected number}
\begin{figure}[t]
  \resizebox{\hsize}{!}{\includegraphics{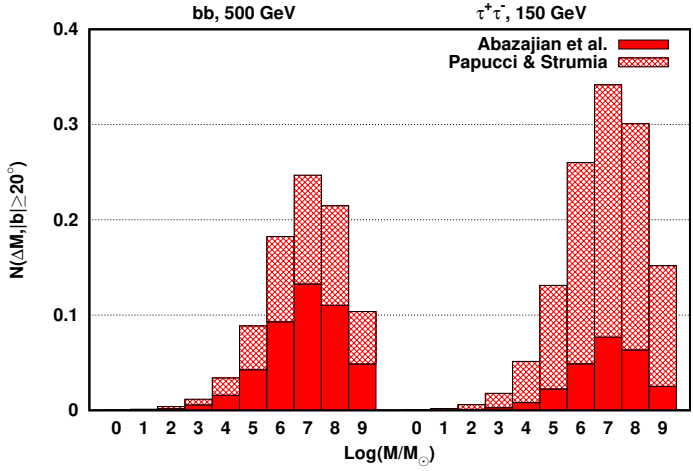}}
  \caption{Expected number of one-year detectable subhaloes per mass
    decade at $|b|\geq20^\circ$. The subhaloes' mass and spatial
    distributions were adopted from Aquarius (Appendix
    \ref{app:sub_diff}) while the SHM concentration was
    used. $M_\mathrm{min} = 10^{-6}\,\mathrm{M}_\odot$. The number
    corresponding to the $b\overline{b},\,\,m_\chi =
    500\,\mathrm{GeV}\,\,(\tau^{+}\tau^{-},\,\,m_\chi =
    150\,\mathrm{GeV})$ WIMP model is shown in the left (right)
    panel. The mass assigned to each bar denotes the geometric mean of
    the interval. Red and red-patterned bars show the number
    considering observational constraints on $\sigmav_\mathrm{eff}$ by
    \citet{2010JCAP...11..041A} and \citet{2010JCAP...03..014P},
    respectively.}
  \label{fig:clump_number}
\end{figure}

Integrating over the mass and spatial distribution
(Eq. \ref{eq:mass_spatial_distr}) reveals the total number of
detectable subhaloes with masses $M_\mathrm{I} \in [M;M+\Delta M]$,
solar distances $D_\mathrm{I} \in [D;D+\Delta D]$, concentrations
$c_\mathrm{I} \in [c;c+\Delta c]$, galactic latitudes $b_\mathrm{I}
\in [b,b + \Delta b]$, and galactic longitudes $l \in [0;2\pi]$, $N =
N(M_\mathrm{I},D_\mathrm{I},c_\mathrm{I},b_\mathrm{I})$,
\begin{eqnarray}
 N & = & \hskip -0.15cm \int\limits_{M_\mathrm{I}} \hskip -0.15cm
 \mathrm{d}M \hskip -0.1cm \int\limits_{D_\mathrm{I}} \hskip -0.1cm
 \mathrm{d}D\,D^2 \hskip -0.1cm \int\limits_{c_\mathrm{I}} \hskip
 -0.1cm \mathrm{d}c \hskip -0.1cm \int\limits_{b_\mathrm{I}} \hskip
 -0.1cm \mathrm{d}b \,\cos b \hskip -0.1cm \int\limits_0^{2\pi} \hskip
 -0.1cm \mathrm{d}l\, P(c,\overline{c})\,
 \frac{\mathrm{d}n_\mathrm{sh}(D_\mathrm{gc},M)}{\mathrm{d}M} \\ 
  & = & \hskip -0.1cm a_N \hskip -0.15cm \int\limits_{M_\mathrm{I}} \hskip
 -0.1cm \mathrm{d}M\, M \hskip -0.15cm
 \int\limits_{0}^{\theta_\mathrm{s}^\mathrm{max}} \hskip -0.15cm
 \mathrm{d}\theta_\mathrm{s}
 \frac{\cos^2\theta_\mathrm{s}}{\sin^4\theta_\mathrm{s}} \hskip -0.6cm
 \int\limits_{\hskip 0.5cm
   c_\mathrm{min}(\theta_\mathrm{s},M)}^{\infty} \hskip -0.65cm
 \mathrm{d}c\,c^{-3} \hskip -0.15cm \int\limits_{b_\mathrm{I}} \hskip
 -0.15cm \mathrm{d}b\,\cos b \hskip -0.15cm
 \int\limits_{0}^{2\pi} \hskip -0.1cm
 \mathrm{d}l\,P(c,\overline{c})\,\frac{\mathrm{d}n_\mathrm{sh}}{\mathrm{d}M},
 \nonumber
\end{eqnarray}
using $D = r_\mathrm{s}/\tan \theta_\mathrm{s}$ (see
Sect. \ref{sect:candidate_sources}) and $a_N \equiv 3/(4\pi
\Delta_\mathrm{c} \rho_\mathrm{crit})$. Parameters defining the
subhalo distribution were taken to match the Aquarius simulation
(Appendix \ref{app:sub_diff}). In total, this resulted in about
$6.4\times 10^{14}$ Galactic subhaloes residing in the Galaxy. For
every single $\theta_\mathrm{s}$ and $M$, the integral counts
detectable subhaloes only, i.e., their concentration is sufficiently
high to ensure their $\sigmav_\mathrm{eff}$ to be smaller than the
observational constraints (cf., Fig. \ref{fig:fiducial}). Therefore,
the lower bound of the concentration integral
$c_\mathrm{min}(\theta_\mathrm{s},M)$ is determined via
Eq. \ref{eq:boost}, choosing the instrument's sensitivity and
constraints from \citet{2010JCAP...11..041A} and
\citet{2010JCAP...03..014P}, respectively. To account for the fact
that highly extended objects will be hardly detectable (see Appendix
\ref{app:mc}), we conservatively chose
$\theta_\mathrm{s}^\mathrm{max}=1^{\circ}$. For the SHM,
Fig. \ref{fig:clump_number} shows the number of detectable subhaloes
per mass decade expected in one year of data taking while considering
subhaloes at galactic latitudes $|b|\geq20^\circ$ only (cf.,
Sect. \ref{sect:1FGL_searches}). Given the dependence of
$c_\mathrm{min}$ on the WIMP model, the results for the
$b\overline{b},\,m_\chi = 500\,\mathrm{GeV}$ model are compared to the
$\tau^+\tau^-,\,m_\chi = 150\,\mathrm{GeV}$ scenario.

Considering the constraints found by \citet{2010JCAP...11..041A} for
the $b\overline{b}\,\, (\tau^{+}\tau^{-})$ model, on average 0.4 (0.2)
subhaloes within $10^5$ and $10^8\,\mathrm{M}_\odot$ are anticipated
for detection with \textit{Fermi}-LAT in one year. Given the Poisson
distribution of $N$, this means that up to \textit{one} massive
subhalo is expected in the one-year data set (at 95\%
confidence). Comparable results have been claimed by other authors,
e.g., \citet{2008JCAP...07..013B}, \citet{2008ApJ...686..262K},
\citet{2008MNRAS.384.1627P,2011PhRvD..83b3518P}, and
\citet{2010ApJ...718..899A}.

In general, note that numerical simulations like Aquarius and Via
Lactea II neglect the influences of baryonic matter distributed in
galactic disks. A recent study by \citet{2010ApJ...709.1138D}
indicates that a baryonic disk may reduce the number of (massive)
subhaloes in the inner galaxy by a factor of 2 to 3.

\subsubsection{Properties}
Given a particular $\gamma$-ray source (such as the fiducial of
Sect. \ref{sect:fiducial_candidate}), massive subhaloes between about
$10^5$ and $10^8\,\mathrm{M}_\odot$ require a minimally enhanced
annihilation cross section $\sigmav_\mathrm{eff}$, see
Fig. \ref{fig:fiducial}. Consistently, the probability for these
objects to appear in current data sets peaks for high subhalo masses
(Fig. \ref{fig:clump_number}). Therefore, subhalos with masses of
$10^5$ up to $10^8\,\mathrm{M}_\odot$ in corresponding distances from
0.5 to 10\,kpc (Fig. \ref{fig:distance}) are favoured for detection
with \textit{Fermi}-LAT as faint and moderately extended
sources. Although for the FHM a DM origin of the fiducial source is
excluded, within the scatter of the more realistic SHM cross sections
required for sources with $\phi(10\!-\!100\,\mathrm{GeV}) \approx
10^{-10}\,\mathrm{cm}^{-2} \mathrm{s}^{-1}$ and angular extents up to
$\sim 1^\circ$ are well consistent with observational constraints. The
presence of sub-substructure will even lower $\sigmav_\mathrm{eff}$
by a mass-dependent factor of \mbox{$\sim 2$} to 3 for massive
subhaloes \citep{2008ApJ...686..262K,2009JCAP...06..014M}. In case of
a cuspier profile ($\gamma=1.2$), the required cross section is
lowered by an additional factor of 1.5. At least for WIMPs of $m_\chi
= 150\,\mathrm{GeV}$ annihilating to $\tau^+\tau^-$, this leads to a
required enhancement of the order of unity within the
scatter. Summarising, in optimistic but realistic scenarios a
$\gamma$-ray emitter at the (one year) detection level of
\textit{Fermi}-LAT with a measured extent $\theta_{68}$ of about
$0.5^\circ$ might be consistent with a subhalo driven by
self-annihilating DM.

In the next years, \textit{Fermi}-LAT will provide deeper observations
with increased observation time $T_\mathrm{obs}$. This will improve
the sensitivity by roughly $\sqrt{T_\mathrm{obs}}$, leading to a
factor of about 2 lower values of the minimum detectable flux for the
five-year catalogue. This in turn will allow us to detect fainter
subhaloes with a correspondingly reduced minimum
$\sigmav_\mathrm{eff}$. The average number of detectable subhaloes
within five years is about 1.3 (0.8) for the
\mbox{$b\overline{b},\,m_\chi = 500\,\mathrm{GeV}$}
\mbox{$(\tau^+\tau^-,\,m_\chi = 150\,\mathrm{GeV})$} scenario.

Via Eq. \ref{eq:boost}, a comparison with observational constraints on
$\sigmav_\mathrm{eff}$ allows to estimate the maximum flux subhalo
candidates are expected to have. The catalogued flux is
$\phi_\mathrm{p} \propto \theta_\mathrm{s}^2/h(\theta_\mathrm{s})$,
which is quadratic for $\theta_\mathrm{s} \ll 0.3^\circ$ and linear in
the limit of large $\theta_\mathrm{s}$. The increase with
$\theta_\mathrm{s}$ originates from decreasing subhalo
distance. Because massive subhaloes require minimum
$\sigmav_\mathrm{eff}$, a subhalo of $10^6\,\mathrm{M}_\odot$ is
assumed below. Within the concentration scatter, the maximum flux
between 10 and 100\,GeV expected for a source with $\theta_\mathrm{s}
= 1^\circ$ is
\begin{equation}
\phi_\mathrm{p}^\mathrm{max}(10\!-\!100\,\mathrm{GeV}) \simeq
1.2\,(2.8)\times
10^{-11}\,\frac{\sigmav_\mathrm{eff}}{\sigmav_0}\,\mathrm{cm}^{-2}\,\mathrm{s}^{-1}
\end{equation}
in the $b\overline{b},\,m_\chi = 500\,\mathrm{GeV}$ and
$\tau^+\tau^-,\,m_\chi = 150\,\mathrm{GeV}$ scenario,
respectively. Given the observational constraints of
\citeauthor{2010JCAP...11..041A}, the high-energy flux of catalogued
candidates should not exceed \mbox{$\phi_\mathrm{p}^\mathrm{max}
  \lesssim 4.0\,(2.8)\times
  10^{-10}\,\mathrm{cm}^{-2}\,\mathrm{s}^{-1}$}.

\section{Searches for DM subhaloes in the 1FGL} \label{sect:1FGL_searches}
In the previous section we demonstrated that DM subhaloes could
appear in $\gamma$-ray catalogues of sufficient sensitivity as faint,
non-variable, and moderately extended objects without astrophysical
counterparts. The 11-month\footnote{August 2008 to July 2009}
point-source catalogue of \textit{Fermi}-LAT
\citep[1FGL,][]{2010ApJS..188..405A} lists 1451 sources significantly
detected above 100\,MeV (test statistic $\mathrm{TS} \geq 25$,
corresponding to a significance $S = 4.1\sigma$), together with the
flux in five discrete energy bins (up to 100\,GeV), position,
significance of variability, and spectral curvature. Source spectra
have been fitted with power laws. Among the sources, 630
objects\footnote{Adding sole associations with other $\gamma$-ray
  catalogues, 671 sources are ``unassociated``.} are not confidently
associated with known sources at other wavelengths.

Although sophisticated methods have been applied to find
multi-wavelength associations for unidentified sources, all algorithms
suffer from lacking sensitivity or incomplete sky-coverage of current
surveys. Therefore, the sample of unassociated high-latitude
\textit{Fermi}-LAT sources is expected to be composed of several
source classes, among them faint AGN (Active Galactic Nuclei), galaxy
clusters, and new exotic sources like DM subhaloes \citep[][and
  references
  therein]{2010MNRAS.408..422S,2010arXiv1007.2644M}. Concerning the
1FGL catalogue, improved association methods recently presented by
\citet{2010arXiv1007.2644M} revealed that $\lesssim 20\%$ of all
unassociated 1FGL sources with $|b| \geq 15^\circ$ may contain new
$\gamma$-ray emitters.

To single out possible subhalo candidates within the sample of
unassociated sources \citep[cf.,][]{2010PhRvD..82f3501B}, we searched
the sample for non-variable\footnote{The cut is passed by sources with
  a steadiness probability $P_\mathrm{s} > 1\%$.}  sources detected
between 10 and 100\,GeV. Requiring a detection at high energy provides
subhalo candidates driven by heavy WIMPs and avoids confusion with
high-energy pulsars\footnote{The spectral properties of $\gamma$-ray
  pulsars can mimic the spectra of DM subhaloes, see
  \citet{2007ApJ...659L.125B}. However, spectral cut-off energies of
  $\gamma$-ray pulsars are well below 5\,GeV
  \citep{2009Sci...325..848A}, excluding a detection above
  10\,GeV.}. Furthermore, the candidate's location was constrained to
galactic latitudes $|b|\geq20^\circ$ to avoid a general confusion
with Galactic sources. Additionally, the lower Galactic foreground
improves the detection sensitivity of \textit{Fermi}-LAT at high
latitudes \citep{2009ApJ...697.1071A}.

Applying all cuts, \textit{twelve} unassociated sources remain. The
twelve sources are listed in Table \ref{tab:clump_candidates} together
with additional information from the catalogue. With the exception of
1FGL~J0614.1-3328, the sample consists of sources at the faint end of
the entire 1FGL sample. Given the result of
\citet{2010arXiv1007.2644M}, the sample should statistically contain
two to three subhaloes at most, consistent with the estimate discussed
previously (Fig. \ref{fig:clump_number}). The expectation of the
sample consisting mostly of AGN is met by applying the same cuts to
all AGN detected by \textit{Fermi}-LAT. A comparison with the sample
of unassociated sources reveals similar cut efficiencies (5\%
vs. 11\%, see Table \ref{tab:efficiencies}), indicating that the two
populations share common properties. Note that for the AGN the
variability cut has subdominant influence as well, see Table
\ref{tab:efficiencies}. Except for three, all AGN that passed the cuts
have been classified as BL Lac.

\begin{table}[t]
 \caption{\label{tab:efficiencies}Cut efficiencies on the sample of
   unassociated sources and AGN.}
 \begin{center}
 \begin{tabular}{lcc}
\hline \hline
Cut & Unassociated & AGN \\
\hline
 -- & 671 & 693 \\
$|b| \geq 20^\circ$ & 249 (100\%) & 539 (100\%) \\
non-variable & 241 (97\%) & 372 (69\%)\\
detected between & \multirow{2}{*}{12 (5\%)} & \multirow{2}{*}{58 (11\%)} \\
$10\!-\!100\,\mathrm{GeV}$ &  & \\
\hline
 \end{tabular}
\tablefoot{The cuts are cumulative, i.e., each number includes all
  cuts listed by previous rows. See text for details.}
 \end{center} 
\end{table}

Even though the twelve candidate objects are listed in the 1FGL
catalogue as unassociated, we extended the counterpart search
to a wider choice of astronomical catalogues. Table
\ref{tab:clump_candidates} lists the classifications of counterpart
candidates in the 68\% confidence regions around the most likely 1FGL
positions, retrieved from the NASA/IPAC Extragalactic Database
(NED). In particular, possible associatons are provided by radio and
X-ray sources, since most of the selected $\gamma$-ray sources are
expected to be AGN. Given that no detailed association study was
conducted, some of the tabulated sources might be by-chance
associations.

\begin{table*}
\caption{\label{tab:clump_candidates}Unassociated, non-variable 1FGL
  sources at high galactic latitudes.}
\begin{center}
 \begin{tabular}{l c c c c c l c}
 \hline \hline
  \multicolumn{1}{l}{Name} & $\sigma_{68}/\sigma_{95}$ & $S$ & $f_\mathrm{p} (0.1\! -\! 100\,\mathrm{GeV})$ & \multirow{2}{*}{$\Gamma$} & $\phi_\mathrm{p} (10\! -\! 100\,\mathrm{GeV})$ & \multicolumn{1}{c}{Possible associations\tablefootmark{a}} & \multicolumn{1}{c}{Remarks}\\
  \multicolumn{1}{c}{1FGL J} & [arcmin] & $[\sigma]$ &  [$10^{-11}\,\mathrm{erg}\,\mathrm{cm}^{-2}\,\mathrm{s}^{-1}$] &  & [$10^{-10}\,\mathrm{cm}^{-2}\,\mathrm{s}^{-1}$] & \multicolumn{1}{c}{68\% c.l.} \\
\hline
 0022.2-1850 & 6.0/9.6 & 9.4 & $1.3(4)$ & $1.6(1)$ & 1.6(7) & RadioSs (4,21,22), Gs ($20 - 18$) & \\
 0030.7+0724 & 3.0/5.1 & 5.8 & $1.0(4)$ & $1.7(4)$ & 1.5(7) & --- & \\
 0051.4-6242 & 2.4/4.2 & 12.0 & $1.8(5)$ & $1.7(1)$ & 1.7(8) & Gs (20), XrayS (3.8) & c\\
 0143.9-5845 & 3.0/4.7 & 9.0 & $1.4(4)$ & $2.0(2)$ & 2.0(9) & RadioS (28$^\S$), Gs (20 - 13) & \\
 0335.5-4501 & 2.4/4.0 & 8.6 & $1.5(4)$ & $2.1(2)$ & 1.6(8) & Gs (19,18) & \\
 0614.1-3328 & 1.2/1.7 & 54.4 & $11.2(6)$ & $1.93(3)$ & 3(1) & GrayS & b\\
 0848.6+0504 & 5.4/8.6 & 5.4 & $1.0(5)$ & $1.2(3)$ & 1.6(8) & RadioSs (2,3,5), Gs \& *s, XrayS (4.4) & c\\
 1323.1+2942 & 1.8/2.7 & 11.9 & $1.5(4)$ & $2.0(1)$ & 2.1(8) & RadioSs (2.8,263,724), Gs \& *s & \\
 1754.3+3212 & 2.4/4.1 & 15.6 & $2.6(4)$ & $2.09(9)$ & 1.4(7) & RadioS (38$^\dag$) & \\
 2134.5-2130 & 3.0/5.1 & 6.7 & $1.1(3)$ & $1.9(2)$ & 1.4(7) & RadioS (22), Gs (20) & \\
 2146.6-1345 & 3.0/4.4 & 9.8 & $1.5(5)$ & $1.8(2)$ & 1.8(8) & RadioS (23), Gs (20), XrayS (1.9) & c\\
 2329.2+3755 & 1.2/1.9 & 10.4 & $1.7(5)$ & $1.6(2)$ & 2.4(9) & G (14) & c\\
\hline
 \end{tabular}
\tablefoot{The columns list the positional uncertainty
  $\sigma_{68\,(95)}$ [68\% (95\%) c.l., semimajor axis], detection
  significance $S$ in Gaussian sigma, integrated energy flux
  $f_\mathrm{p} (0.1-100\,\mathrm{GeV})$, spectral index $\Gamma$, and
  the photon flux $\phi_\mathrm{p}(10\!-\!100\,\mathrm{GeV})$. Here,
  parentheses indicate the corresponding error on the last
  decimal(s). Furthermore, the type classifications of sources found
  in astronomical catalogues within the 68\% uncertainty region of the
  \textit{Fermi}-LAT position are listed.
  \tablefoottext{a}{Classifications referred to are RadioS (radio
    source), G (galaxy), * (star), XrayS (X-ray source), and GrayS
    ($\gamma$-ray source). For radio, optical, and X-ray sources
    corresponding fluxes are given in mJy (at 1.4\,GHz [$^{(\S)}$:
      843\,MHz, $^{(\dag)}$: 4.85\,GHz]), apparent magnitudes, and
    $10^{-12}\,\mathrm{erg}\,\mathrm{cm}^{-2}\,\mathrm{s}^{-1}$,
    respectively. The unabsorbed X-ray flux was derived from the
    catalogued count-rate, assuming a power law with index 2.0 (with
    WebPIMMS, http://heasarc.gsfc.nasa.gov/Tools/w3pimms.html). The
    hydrogen column density was obtained from the LAB survey, see
    Sect. \ref{sect:swift_data}. Sources referred to are listed in the
    FIRST \citep{1995ApJ...450..559B}, JVAS/CLASS
    \citep{2007MNRAS.376..371J}, NVSS, SUMSS
    \citep{2003MNRAS.342.1117M}, 2MASS, APMUKS
    \citep{1990MNRAS.243..692M}, SDSS \citep{2009ApJS..182..543A},
    ROSAT, or EGRET \citep{1999ApJS..123...79H} catalogue,
    respectively.}  \tablefoottext{b}{The spectrum is probably
    curved.}  \tablefoottext{c}{The $\gamma$-ray source has
    been associated by a cross-correlation of unidentified
    \textit{Fermi}-LAT sources with the ROSAT All Sky Survey Bright
    Source Catalogue (see \citet{2010MNRAS.408..422S} for details).}}
 \end{center}
\end{table*}

Governed by lacking association, faintness, and spectral shape, this
study focusses on the most promising candidate, \clumpi. Within the
errors, its high-energy flux and spectral index are well-compatible
with a self-annihilating DM scenario. The source has only been
detected between 10 and 100\,GeV, see Fig. \ref{fig:spectra}.

\subsection{Multi-wavelength properties of \clumpi}
\subsubsection{Catalogued data} \label{sect:catalogued_data}
\textit{No} counterpart candidate was found within the positional
uncertainty of the $\gamma$-ray source at a 68\% confidence level (Table
\ref{tab:clump_candidates}). In the 95\% confidence region, the faint
radio object \mbox{NVSS J003030+072132} is located
\citep[$f_{1.4\,\mathrm{GHz}} = (3.5 \pm
  0.4)\,\mathrm{mJy}$;][]{1998AJ....115.1693C}. However, no conclusive
infrared \citep[2MASS,][]{2006AJ....131.1163S} or optical \citep[USNO
  B1.0,][]{2003AJ....125..984M} association of the NVSS source is
known so far\footnote{Within the 2$\sigma$ positional uncertainty of
  \mbox{NVSS J003030+072132}, a very faint optical SDSS (Sloan Digital
  Sky Survey) source is located ($26.0^\mathrm{m}$) -- SDSS
  J003031.22+072132.2 \citep[SDSS DR7,][]{2009ApJS..182..543A}. 
  However, this object was observed 
  with the edge of the plated SDSS camera. Therefore, this detection
  is probably spurious.}. Note that no dSph galaxy is located in the
source region (NED). ROSAT \citep[$0.1 -
  2.4\,\mathrm{keV}$;][]{1999A&A...349..389V} observations of the
region with an exposure of about 170\,s revealed no X-ray source down
to an energy-flux level of \mbox{$\sim
  10^{-12}\,\mathrm{erg}\,\mathrm{cm}^{-2}\,\mathrm{s}^{-1}$}
\citep{2010Borm}.

We emphasise that the orphaned faint radio source is likely located in
the uncertainty region of \clumpi\space by chance, because about 0.7
NVSS sources are expected by statistics.

\begin{figure}[t]
  \resizebox{\hsize}{!}{\includegraphics{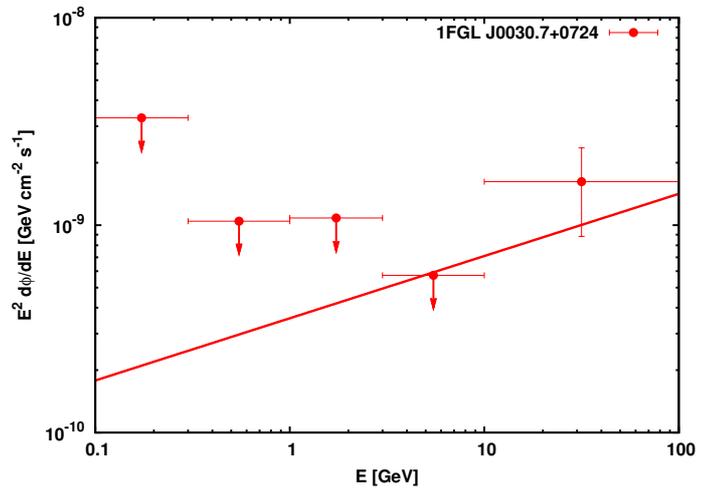}}
  \caption{Energy spectrum of \clumpi\space as given in the first
    \textit{Fermi}-LAT catalogue between 100\,MeV and 100\,GeV (filled
    circles). The solid red line depicts the (catalogued) power law
    fitting the data. Only the highest energy bin has been
    significantly detected, while for the low-energy bins upper limits
    are shown (95\% c.l.).}
  \label{fig:spectra}
\end{figure}

\subsubsection{\textit{Fermi}-LAT data} \label{sect:fermi_data}
By analysing the 24-month public archival data between 10 and
100\,GeV, updated results on \clumpi\space will be provided. For the
same energy range, a reanalysis of the 11-month data is presented for
comparison. Particular focus will be drawn on positional properties,
the high-energy flux, and the photon distribution, which allows us to
investigate possible counterparts, temporal variability, and the
angular extent.

The data analysis was performed with the latest public version of the
\textit{Fermi Science Tools} (v9r18p6)\footnote{Fermi Science Support
  Center, http://fermi.gsfc.nasa.gov/ssc/} along with recommended
options and the set of instrument-response functions
\textit{P6\_V3\_DIFFUSE} \citep{2009arXiv0907.0626R}. Throughout the
analysis, the optimiser MINUIT was used. For reliable results, photons
of event class 3 (\textit{Diffuse}) and 4 (\textit{DataClean}) within
a radius of 10$^\circ$, centred on the nominal position of \clumpi,
were selected. Given that most of the photons are at highest energies
(Fig. \ref{fig:spectra}), only events between 10 and 100\,GeV were
selected to minimise the background and ensure a narrow PSF. The data
were processed using \textit{gtselect}, \textit{gtmktime},
\textit{gtltcube}, \textit{gtexpmap}, and unbinned \textit{gtlike}. To
compute the most likely position and its corresponding uncertainty on
basis of the $10\!-\!100$\,GeV photon sample, we used
\textit{gtfindsrc}. For the purpose of detailed counterpart searches,
the two-dimensional likelihood function $L(\mathrm{R.A.,Dec.})$ was
computed, which provides the 95\% uncertainty contour by
$2\,\Delta(\log L) = 6.18$ (2 degrees of freedom).

\begin{table}[t]
 \caption{\label{tab:analysis_plfit} Positional and spectral
   properties of \clumpi\space as given in the catalogue (11\,months)
   and by the (re-)analysis of the first 11 and 24-month data sets.}
 \begin{center}
 \begin{tabular}{l c c c c c c}
\hline \hline
 \hspace{-2mm} Data & \hspace{-3mm} $E$ & \hspace{-2mm} R.A. & \hspace{-2mm} Dec. & \hspace{-4mm} $\phi_\mathrm{p}(10\!-\!100\,\mathrm{GeV})$ & \hspace{-3mm} $S(\phi_\mathrm{p})$ \hspace{-3mm}\\
 \hspace{-2mm} set & \hspace{-3mm} [GeV] & \hspace{-2mm} (J2000) & \hspace{-2mm} (J2000) & \hspace{-4mm} [$10^{-10}\,\mathrm{cm}^{-2}\,\mathrm{s}^{-1}$] &\hspace{-6mm} [$\sigma$] \hspace{-3mm} \\
\hline
  \hspace{-2mm} 1FGL & \hspace{-3mm} 0.1--100 & \hspace{-2mm} 00 30 42.6 & \hspace{-2mm} +07 24 09 & \hspace{-4mm} $1.5 \pm 0.7$ &\hspace{-3mm} 6.6 \hspace{-3mm}\\
  \hspace{-2mm} 11 & \hspace{-3mm} 10--100 & \hspace{-2mm} 00 30 37.6 & \hspace{-2mm} +07 24 15 & \hspace{-4mm} $1.4 \pm 0.7$ &\hspace{-3mm} 6.5 \hspace{-3mm}\\
  \hspace{-2mm} 24 & \hspace{-3mm} 10--100 & \hspace{-2mm} 00 30 47.6 & \hspace{-2mm} +07 24 20 & \hspace{-4mm} $0.9 \pm 0.4$ &\hspace{-3mm} 6.6 \hspace{-3mm}\\
\hline
 \end{tabular}
\tablefoot{The second column lists the analysed energy range. The
  11\,(24)-month analysis focusses on the high-energy flux
  $\phi_\mathrm{p}(10\!-\!100\,\mathrm{GeV})$ only. In all cases, the
  significance $S$ of the high-energy bin is well above 6$\sigma$.}
 \end{center} 
\end{table}

The source model for the data analysis contains all 1FGL sources
within the region of interest (ROI, radius $10^\circ$). Their
parameters were taken as catalogued and we used the latest Galactic
(\textit{gll\_iem\_v02.fit}) and extragalactic
(\textit{isotropic\_iem\_v02.txt}) diffuse background models. All
parameters but those of \clumpi\space were kept fixed. Furthermore,
the catalogued power-law index of \clumpi\space was used while fitting
the flux between 10 and 100\,GeV. Although the exposure of the
24-month data has almost doubled with respect to the catalogue, the
use of the catalogued properties for sources within the ROI will not
affect the analysis between 10 and 100\,GeV: The three nearby sources,
i.e., 1FGL~J0022.5+0607, 1FGL~J0030.4+0451, and 1FGL~J0023.5+0930, are
not only more than 2$^\circ$ away from \clumpi, but they are also not
significantly detected between 10 and 100\,GeV. Furthermore, visual
inspection does not reveal any other relevant source within this
nearby region.

\begin{table}[t]
 \caption{\label{tab:photons}High-energy photons detected from
  \clumpi\space within $0.5^\circ$.}
 \begin{center}
 \begin{tabular}{l c c c c c c}
\hline \hline
 $E$ & R.A.& Dec. & $\vartheta$ & $\Delta t$ & \multirow{2}{*}{CT} & event\\
 $[$GeV$]$ & [deg] & [deg] & [deg] & [30\,d] &  & class\\
\hline
83.8 & 7.6330 & 7.3975 & 56.26 & 2.46 & B & 3\\
11.8 & 7.7293 & 7.3771 & 36.39 & 5.19 & F & 4\\
39.8 & 7.7841 & 7.4962 & 47.38 & 7.96 & B & 4\\
10.2 & 7.6426 & 7.4483 & 34.21 & 10.46 & F & 4\\
15.0 & 7.6361 & 7.1872 & 38.24 & 11.12 & B & 4\\
43.8 & 7.8392 & 7.4151 & 20.81 & 18.93 & F & 4\\
\hline
 \end{tabular}
\tablefoot{The table lists their energy $E$, celestial position
  (J2000), inclination $\vartheta$, detection time $\Delta t$, and
  conversion type (CT). By $\Delta t$ the time between detection and
  mission start is given. The conversion type is front (F) or back
  (B). For each event, we list the classification assigned by LAT data
  reconstruction (Pass 6), where 3 tags the \textit{Diffuse} and 4 the
  \textit{DataClean} class \citep[see][]{2010PhRvL.104j1101A}.}
 \end{center} 
\end{table}

The analysis of the 11-month data reproduces the catalogued values
well (Table \ref{tab:analysis_plfit}). After 11\,(24)\,months, five
(six) photons between 10 and 100\,GeV have been detected within a
radius of 0.5$^\circ$ around the nominal position, listed in Table
\ref{tab:photons}. Except one, all photons are classified as class 4
events and are therefore very likely signal events. The Galactic
foreground and the extragalactic background at the source
position are negligible with respect to the signal, with an expected
total number of background photons $N_\mathrm{bg} = 0.6\,(1.2)$ within
the considered region of $0.5^\circ$. For comparison, the predicted
number of signal events is $N_\mathrm{sig} = 4.9\,(5.8)$ after
11\,(24)\,months. According to the 11-month data set, the
(10--100\,GeV) best-fit position shifts by about $2.5^\prime$. The
small positional error of the sixth photon also accounts for the
increase of the source's positional uncertainty, see
Fig. \ref{fig:skyplot}.

The average flux over the entire data set has decreased by a factor of
roughly 1.5 with respect to the first 11\,months (Table
\ref{tab:analysis_plfit}). To judge on the variability of \clumpi, its
temporal photon distribution (Table \ref{tab:photons}) was tested
for compatibility with a constant flux, using an unbinned
Kolmogorov-Smirnov (KS) test \citep{2007NR}. The KS test is already
valid for low photon counts, unlike the binned chi-square method used
by the catalogue. The KS test confirms the null-hypothesis of
non-variability with a probability of about 0.7\,(0.5) for the
11\,(24) months data set. The varying exposure on the region was
taken into account by examining the photon distribution of the bright
pulsar nearby (1FGL~J0030.4+0451).

The analysis of the (intrinsic) spatial extent of the source is based
upon a likelihood-ratio test, using all photons listed in Table
\ref{tab:photons}. The corresponding statistical measure is given by
\mbox{$L(\theta_\mathrm{s}) = -2 \sum_{i=1}^N \ln
  [p_\mathrm{det}(\mathbi{x}_i-\overline\mathbi{x};\theta_\mathrm{s})
    + b]$}, where $p_\mathrm{det}(\mathbi{x};\theta_\mathrm{s})$ is
the probability distribution function for a photon detected at
$\mathbi{x}$, $\overline\mathbi{x}$ denotes the best-fit position
(Table \ref{tab:analysis_plfit}), and $b$ incorporates the flat
background. For a spatially extended $\gamma$-ray emitter
$p_\mathrm{det} = p_\mathrm{PSF} \ast p_\mathrm{int}$, the (two
dimensional) convolution of the \textit{Fermi}-LAT PSF (P6\_v3,
diffuse class) with the intensity profile of the emitter. In the
subhalo case, the intensity profile follows the line-of-sight integral
of the squared NFW profile (Eq. \ref{eq:sub_profile}). The quantity
$\Delta L = L - L_\mathrm{min}$ follows a chi-square distribution with
one degree of freedom, with additional terms of the order of
$1/N^{1/2}$, which are important for a small number of counts
\citep{1938Wilks,1979ApJ...228..939C}. The likelihood is minimised
($L_\mathrm{min}$) for the intrinsic extension parameter fitting the
photon distribution best. Examining the 11-month data, the test shows
the source to be consistent with a point source, implying that the
intrinsic extent is smaller than the (average) PSF (about
$0.15^\circ$). The 24-month data favour a moderate extent
$\theta_\mathrm{s} = 0.14^{+0.20}_{-0.12}\,\mathrm{deg}$, which is,
however, not significant. Upper limits on the extension parameter are
$\theta_\mathrm{s} \leq 0.54\, (0.72)\,\mathrm{deg}$ at 95\%
confidence level, derived from the 11\,(24)\,months data set. Since
the low statistics affect the chi-square distribution, note that the
confidence level is not precisely defined
\citep{1979ApJ...228..939C,1996ApJ...461..396M}. Furthermore, we point
out that the PSF of \textit{Fermi}-LAT (P6\_v3) may be
underestimated\footnote{http://fermi.gsfc.nasa.gov/ssc/data/analysis/LAT\_caveats.html}
and changes will have an impact on the fitted extent.

\subsubsection{\textit{Swift}-XRT data} \label{sect:swift_data}
The field was successfully proposed for observation with the X-ray
telescope (XRT, 0.2--10\,keV) aboard the \textit{Swift} satellite
\citep{2004ApJ...611.1005G,2005ApJ...621..558G}. The observations
(Obs.~ID~00041265001) were carried out on 10 November, 2010, between
00:23:46 and 19:52:56\,UT with a total effective exposure of
10.1\,ks. Observations with the XRT were performed in photon-counting
(PC) mode. The XRT data were calibrated and selected with standard
screening criteria (\textit{xrtpipeline}), using the HEAsoft 6.10
package for data reduction with the current version of calibration
files available (release 2010-09-30). For the analysis, events with
grades 0--12 \citep{2005SSRv..120..165B} were used. The spectral
analysis was carried out with \textit{Xspec}
\citep[12.6.0,][]{1996ASPC..101...17A}, using the PC grade 0--12
response matrix \textit{swxpc0to12s6\_20070901v011.rmf} with the
ancilliary response function generated by \textit{xrtmkarf} for PSF
correction and the position of the source considered. The on-source
region was selected to contain about 90\% of the PSF ($\approx
47''$). For background subtraction, an off-source region with radius
of about $4^\prime$ was used. To ensure a spectral fit of
sufficient quality, the spectra were rebinned to a minimum of 10
events per bin (with \textit{grppha}). Owing to the low statistics
accumulated, the C-statistic was used for spectral fitting.

\begin{figure}[t]
  \resizebox{\hsize}{!}{\includegraphics{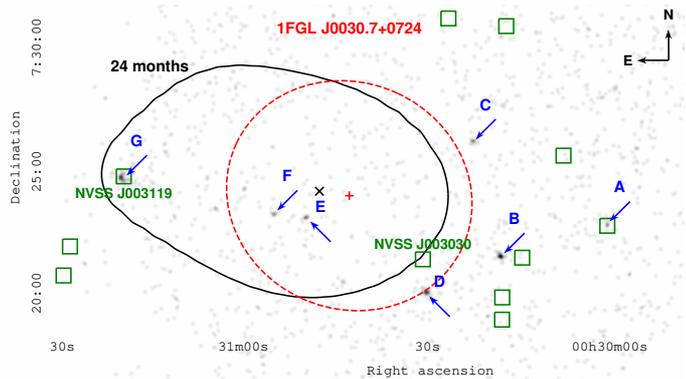}}
  \caption{Celestial region of \clumpi, $25^\prime \times
    13^\prime$. The catalogued position is indicated by the red $+$,
    while the dashed red ellipse borders its uncertainty (95\%
    c.l.). The black $\times$ marks the 24-month position, the solid
    black contour its uncertainty at 95\% confidence. Photons detected
    by \textit{Swift}-XRT (10.1\,ks) are mapped by the back-image,
    which is smoothed with a Gaussian ($7''$). The positions of NVSS
    radio sources are given by the darkgreen boxes, the two NVSS sources
    discussed in the text are named. In this region, seven X-ray
    sources have been discovered, indicated by the blue arrows (see
    Table \ref{tab:swift_srcs}). Note that the boxes' size is chosen
    arbitrarily.}
  \label{fig:skyplot}
\end{figure}

In the field-of-view (FoV) of XRT, seven new X-ray sources were
discovered with a probability of being background fluctuations smaller
than $10^{-6}$. We show them in Fig. \ref{fig:skyplot}. Their
positional properties, measured flux, and the flux corrected for
photoelectric absorption between 0.2 and 2\,keV are listed in Table
\ref{tab:swift_srcs}. The spectra of the two brightest sources are
well-fitted by an absorption corrected power-law model, fixing the
hydrogen column density $N_\mathrm{H}$ to the nominal Galactic
value. The power-law index for the faint sources was fixed to 2.0. The
Galactic hydrogen column density was obtained from the LAB HI survey
\citep{2005A&A...440..775K} for the corresponding celestial positions.

\paragraph{Associations.}
The newly discovered X-ray sources were studied for possible
associations in other accessible wavelengths. Multi-wavelength surveys
covering the region are the NVSS in the radio, the 2MASS in the
infrared, and the USNO B1.0 and SDSS DR7 catalogues for the optical
band. For every \textit{Swift} source we found at least one SDSS
source to be positionally coincident (Table \ref{tab:swift_srcs_cp}),
with apparent magnitudes between 21$^\mathrm{m}$ and
17$^\mathrm{m}$. Owing to insufficient sensitivity, the very faint
SDSS sources have not been detected by USNO.

\begin{table*}[t]
\caption{\label{tab:swift_srcs}X-ray sources detected with the
  \textit{Swift}-XRT.}
\begin{center}
\begin{tabular}{l c c c c c c c c}
\hline \hline
\multirow{2}{*}{ID} & \multicolumn{1}{c}{Name} & $\sigma_\mathrm{90}$ & \multirow{2}{*}{$S/N$} & $f^\mathrm{abs}(0.2\!-\!2\,\mathrm{keV})$ & $N_\mathrm{H}$ & $\phi_0$ & \multirow{2}{*}{$\Gamma$} & $f^\mathrm{unabs}(0.2\!-\!2\,\mathrm{keV})$\\
 & \multicolumn{1}{c}{SWIFT J} & [arcsec] &  & [$10^{-14}$\,erg\,cm$^{-2}$\,s$^{-1}$] & $[10^{20}\,\mathrm{cm}^{-2}]$ & [$10^{-5}$\,keV$^{-1}$\,cm$^{-2}$\,s$^{-1}$] &  & [$10^{-14}$\,erg\,cm$^{-2}$\,s$^{-1}$] \\
\hline
A & 003000.3+072301\tablefootmark{a} & 6 & 3.5 & $3.5_{-0.9}^{+1.1}$ & 3.98 & $1.4 \pm 0.4$ & 2.0 & $5.2 \pm 1.5$\\
B & 003017.8+072142 & 5 & 5.4 & $5.0_{-2.1}^{+3.0}$ & 3.71 & $2.2_{-0.4}^{+0.6}$ & $1.4 \pm 0.3$ & $6.7_{-1.8}^{+2.3}$\\
C & 003022.1+072623\tablefootmark{a} & 6 & 3.1 & $1.7_{-0.4}^{+0.5}$ & 3.10 & $0.6_{-0.2}^{+0.3}$ & 2.0 & $2.2_{-0.7}^{+1.1}$ \\
D & 003030.0+072013\tablefootmark{a} & 5 & 5.1 & $5.2_{-2.0}^{+2.8}$ & 3.71 & $2.0_{-0.4}^{+0.6}$ & 2.0 & $7.4_{-1.5}^{+2.2}$\\
E & 003049.8+072316\tablefootmark{a} & 6 & 3.0 & $3.1_{-1.1}^{+1.0}$ & 3.10 & $1.2_{-0.3}^{+0.4}$ & 2.0 & $4.4_{-1.1}^{+1.5}$ \\
F & 003054.9+072328\tablefootmark{a} & 6 & 2.8 & $2.0_{-0.6}^{+0.8}$ & 3.10 & $0.8_{-0.3}^{+0.4}$ & 2.0 & $3.0_{-1.1}^{+1.5}$ \\
G & 003119.8+072454 & 5 & 6.5 & $15.9_{-5.0}^{+4.5}$ & 3.10 & $6.5_{-0.9}^{+1.1}$ & $1.6 \pm 0.3$ & $20.7^{+8.8}_{-4.7}$\\
\hline
 \end{tabular}
\tablefoot{The FoV is centred on \mbox{$(\mathrm{R.A.},\mathrm{Dec.}) =
  (7.6315,7.4211)\,\mathrm{deg}$} with a radius of $13^\prime$. We give
  an internal ID, the position (SWIFT~JHHMMSS.s$\pm$DDMMSS) and its
  corresponding error at 90\% confidence level $\sigma_{90}$
  (determined with \textit{xrtcentroid}), and the signal-to-noise
  ratio $S/N$ (\textit{Ximage}) of the observed flux
  $f^\mathrm{abs}$. If constraining, a power-law model corrected for
  photoelectric absorption was fitted to the spectrum,
  $\mathrm{d}\phi/\mathrm{d}E =
  \phi_0\,(E/\mathrm{keV})^{-\Gamma}$. The hydrogen column density
  $N_\mathrm{H}$ was fixed during the fit. The unabsorbed flux
  $f^\mathrm{unabs}$ was derived from the power-law fit.
  \tablefoottext{a}{Due to a low S/N a two-parameter power-law fit is
    not constraining. The fluxes were derived assuming the index
    $\Gamma = 2.0$.}}
 \end{center} 
\end{table*}

\begin{table*}[t]
\caption{\label{tab:swift_srcs_cp}Likely counterparts of the
  X-ray sources listed in Table \ref{tab:swift_srcs}.}
\begin{center}
\begin{tabular}{l c c c c c c c c c}
\hline \hline
\multicolumn{1}{l}{ID} & \multicolumn{2}{c}{NVSS\tablefoottext{a}} & \multicolumn{2}{c}{2MASS} & \multicolumn{2}{c}{USNO B1.0\tablefoottext{b}} & \multicolumn{3}{c}{SDSS DR7}\\
 & Name & $S$ [mJy] & Name & K & Name & R & Name & g & Type \\
\hline
A & 003000+072255 & $47(2)$ & -- & -- & 0973-0005315 & 20.0$^\mathrm{m}$& J003000.24+072254.7 & 20.3$^\mathrm{m}$ & 6\\
B & -- & -- & -- & -- & 0973-0005428 & 20.4$^\mathrm{m}$ & J003017.75+072140.6 & 19.6$^\mathrm{m}$ & 6\\
C & -- & -- & -- & -- & -- & -- & J003022.22+072621.3 & 21.4$^\mathrm{m}$ & 6\\
\multirow{2}{*}{D} & \multirow{2}{*}{--} & \multirow{2}{*}{--} & \multirow{2}{*}{00302977+0720101} & \multirow{2}{*}{15.3$^\mathrm{m}$} & 0973-0005481 &  18.9$^\mathrm{m}$ & \multirow{2}{*}{J003029.77+072010.3} & \multirow{2}{*}{18.5$^\mathrm{m}$} & \multirow{2}{*}{3} \\
     & & & & & 0973-0005484 & 19.3$^\mathrm{m}$ &\\
E & -- & -- & -- & -- & -- & -- & J003049.61+072313.5 & 21.0$^\mathrm{m}$ & 6\\
\multirow{2}{*}{F} & \multirow{2}{*}{--} & \multirow{2}{*}{--} & \multirow{2}{*}{00305500+0723233} & \multirow{2}{*}{15.7$^\mathrm{m}$} & \multirow{2}{*}{0973-0005560} & \multirow{2}{*}{18.2$^\mathrm{m}$} & J003054.80+072323.1 & 20.7$^\mathrm{m}$ & 6\\
     & & & & & & & J003055.00+072323.2 & 18.5$^\mathrm{m}$ & 6\\
G & 003119+072456 & $11.6(6)$ & -- & -- & 0974-0005617 & 18.6$^\mathrm{m}$ & J003119.71+072453.5 & 17.4$^\mathrm{m}$ & 6\\
\hline
 \end{tabular}
\tablefoot{Scans ranging from radio (NVSS) to infrared (2MASS) and
  optical (USNO and SDSS) wavelength bands are given. The table lists
  the object's name and the catalogued flux or apparent
  magnitude. Here, parentheses indicate the corresponding error on the
  last decimal. Based on photometric morphology, SDSS provides a
  separation between galaxy-like (3) and star-like objects (6), see
  \citet{2001ASPC..238..269L}.
  \tablefoottext{a}{Frequency $\nu = 1.4\,\mathrm{GHz}$}
  \tablefoottext{b}{The column lists R2. If not
    available, R1 or B1 is given instead \citep[see][and references
      therein]{2003AJ....125..984M}.}  }
 \end{center} 
\end{table*}

\section{Discussion} \label{sect:discussion}
\subsection{An AGN origin of \clumpi\space} \label{sect:HBL_scenario}
\begin{figure}[t]
  \resizebox{\hsize}{!}{\includegraphics{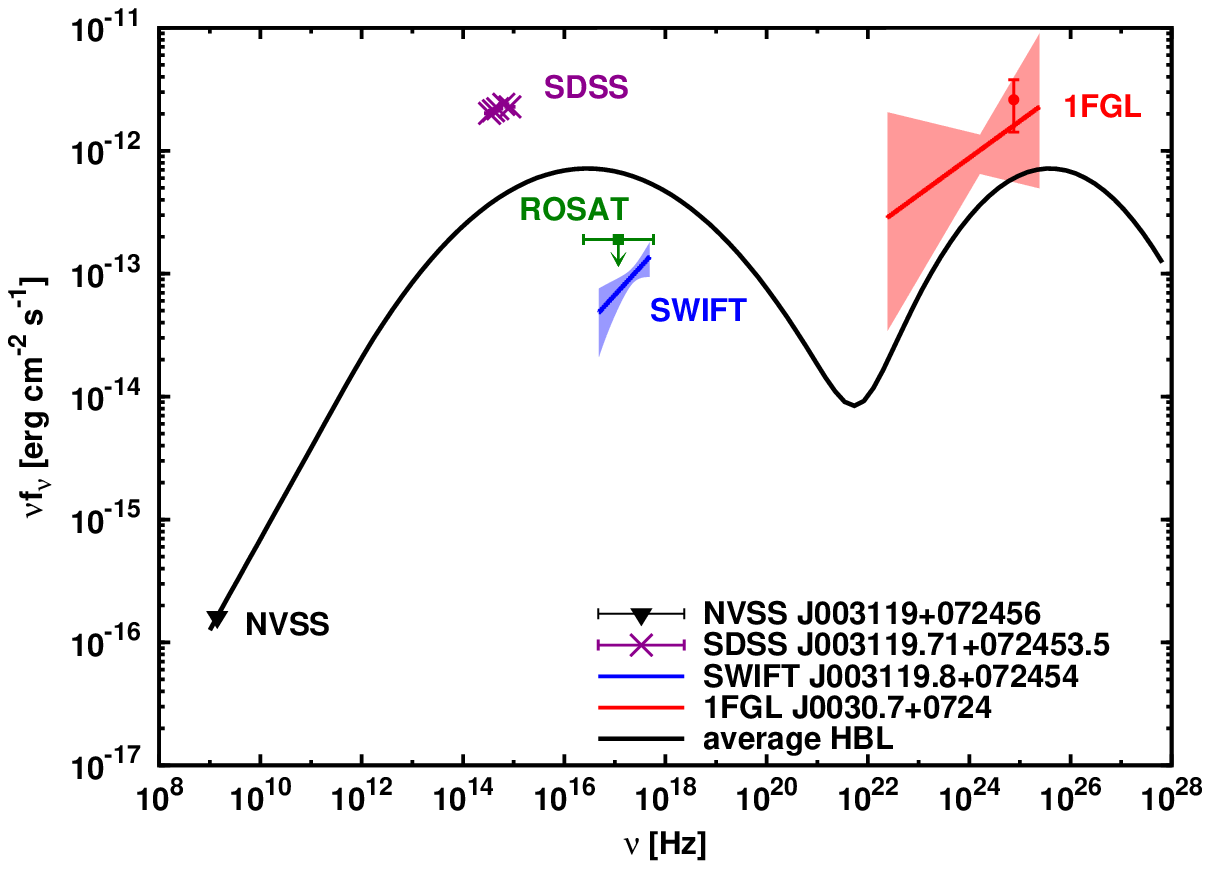}}
  \caption{Energy spectra of \clumpi\space (solid red line) and the
    favoured radio (filled black triangle), optical (violet
    crosses, dereddened $ugriz$ magnitudes), and X-ray (solid blue
    line) counterparts, together with the SED of an average HBL (solid
    black line). The SED was adapted from
    \citeauthor{2001A&A...375..739D}, assuming the average redshift of
    known HBLs $z=0.25$ \citep{2001A&A...375..739D}, and is normalised
    to the radio flux of NVSS~J003119+072456. The frequency-dependent
    energy flux $\nu f_\nu$ is given in the observer's frame. Note that
    the statistical errors of the radio and optical data points are
    too low to be resolved in the figure. Statistical uncertainties of
    the X- and $\gamma$-ray spectra are indicated by the corresponding
    shaded areas, which we derived with Eq. 1 in
    \citet{2009ApJ...707.1310A}. The filled red circle indicates the
    catalogued high-energy flux from \textit{Fermi}-LAT. Observations
    with ROSAT provide an upper limit on the X-ray flux at the nominal
    position of \clumpi\space between 0.1 and 2.4\,keV, which is
    depicted by the green square \citep[95\% c.l., assuming $\Gamma =
      2.0$,][]{2010Borm}.}
  \label{fig:sed_clump}
\end{figure}

The $\gamma$-ray signal of \clumpi\space can be explained by a
conventional AGN. With respect to the unified scheme for the spectral
energy distribution (SED) of AGN (namely Flat Spectrum Radio Quasars
(FSRQs) and blazars), see, e.g., \citet{2001A&A...375..739D}, the hard
spectral index of \clumpi\space ($\Gamma \approx 1.7$) is compatible
with a high-energy-peaked blazar (HBL). Within the updated positional
uncertainty of \clumpi\space derived from the 24-month data
(Fig. \ref{fig:skyplot}), the most likely radio counterpart is
\object{NVSS~J003119+072456} ($f_{1.4\,\mathrm{GHz}} =
(11.6\pm0.6$)\,mJy), which positionally coincides with the newly
discovered hard X-ray source SWIFT~J003119.8+072454 ($\Gamma \approx
1.6$). Note that corresponding to the notation of Table
\ref{tab:swift_srcs}, the \textit{Swift} source is flagged with a
\textit{G} in Fig. \ref{fig:skyplot}. The energy flux observed between
0.2 and 2\,keV is \mbox{$\sim 2\times
  10^{-13}\,\mathrm{erg}\,\mathrm{cm}^{-2}\,\mathrm{s}^{-1}$} (Table
\ref{tab:swift_srcs}). Additionally, an optical counterpart of the
radio and X-ray source is listed in the SDSS catalogue ($r =
17.4^\mathrm{m}$), see Table \ref{tab:swift_srcs_cp}. In
Fig. \ref{fig:sed_clump} we show an empirical model for the average
SED of HBLs, which is based on the bolometric luminosity distribution
of FSRQs and blazars
\citep{1997MNRAS.289..136F,1998MNRAS.299..433F,2001A&A...375..739D}. The
SED is normalised to the radio flux of NVSS~J003119+072456 (at 5
GHz). For comparison, the spectral measurements of the optical and
X-ray counterparts are presented as well. Within the observational
errors and assuming temporal variability, the $\gamma$-ray spectrum of
\clumpi\space is consistent with the model prediction. Furthermore,
the spectral index of the X-ray source agrees with an HBL, while its
flux is fainter than predicted for an (average) HBL. This might be
also explainable by temporal variability (the radio, X-ray, and
$\gamma$-ray observations were not taken simultaneously), and blazars
are well known to be variable in all wavelength bands, where the
amplitude of variability increases with energy
\citep{1997ARA&A..35..445U}.

The other fainter objects in the uncertainty region (the radio source
NVSS~J003030+072132 and the two X-ray sources \textit{E} and
\textit{F}, see Fig. \ref{fig:skyplot}) are less likely to be
associated with \clumpi, but cannot be excluded. For
NVSS~J003030+072132, no X-ray association was detected with
\textit{Swift}-XRT at the level of $2\times
10^{-14}\,\mathrm{erg}\,\mathrm{cm}^{-2}\,\mathrm{s}^{-1}$. No
conclusive optical counterpart is catalogued (above \mbox{$\sim
  26^\mathrm{m}$}, see Sect. \ref{sect:catalogued_data}). With respect
to the comparatively high $\gamma$-ray signal (cf.,
Fig. \ref{fig:sed_clump}), this source therefore fails to provide a
convincing counterpart for \clumpi. Similarly, the lacking radio
detection as well as energy fluxes ($\sim 4\times
10^{-14}\,\mathrm{erg}\,\mathrm{cm}^{-2}\,\mathrm{s}^{-1}$), which are
much fainter than the HBL prediction, disfavour a coincidence of the
X-ray sources \textit{E} and \textit{F} with \clumpi.

\subsection{A DM subhalo origin of \clumpi\space} \label{sect:subhalo_scenario}
Without a clear indication for variability, it remains plausible that
the $\gamma$-ray emission of \clumpi\space originates from a DM
subhalo. The analysis of the arrival times of the source photons
(Sect. \ref{sect:fermi_data}) is consistent with a temporally constant
source of moderate spatial extent. The reconstructed high-energy flux
within the statistical errors is
$\phi_\mathrm{p}(10\!-\!100\,\mathrm{GeV}) \gtrsim 5\times
10^{-11}\,\mathrm{cm}^{-2}\,\mathrm{s}^{-1}$, while the upper limit on
the extent is $\theta_\mathrm{s} \lesssim 0.7^\circ$, corresponding to
$\theta_{68} \lesssim 0.3^\circ$. As shown in
Sect. \ref{sect:Fermi_clumps}, in realistic WIMP scenarios the high
effective self-annihilation cross section required to explain the
source with DM annihilation in a FHM subhalo is hardly compatible with
current observational constraints (see Fig. \ref{fig:fiducial} and
Table \ref{tab:sigmav_eff}). However, given the more realistic SHM,
flux and extent of \clumpi\space are consistent with a subhalo of mass
between $10^6$ and $10^8\,\mathrm{M}_\odot$. Assuming a DM subhalo of
$10^6\,\mathrm{M}_\odot$, the resulting distance
would be $2.4_{-0.7}^{+1.0}\,\mathrm{kpc}$, given the concentration
scatter of the SHM model. For a WIMP of 500\,GeV annihilating to
$b\overline{b}$, the required minimum effective enhancement is 7 for a
high-concentrated SHM subhalo with a corresponding distance of
$1.7\,\mathrm{kpc}$, while it increases to 31 for an
average-concentrated subhalo with a corresponding distance of
$2.4\,\mathrm{kpc}$. Note that $h(0.7^\circ) \approx
1.4$. An even lower boost factor is required for a lighter WIMP of
150\,GeV which predominantly annihilates to $\tau^+\tau^-$: 3\,(13)
for a high-concentrated (average-concentrated) SHM subhalo. Further
decrease of the necessary boost may be provided by sub-substructure
and cuspier profiles (Sect. \ref{sect:Fermi_clumps}).

In addition to theoretical uncertainties on halo properties and their
expected scatter (Sect. \ref{sect:density}), observational
uncertainties affect the distance and boost factor estimates. The
uncertainties on the flux directly change the boost, while
uncertainties on the most crucial measurement, the angular extent
$\theta_\mathrm{s}$, affect both the required boost and the
distance estimate (Sect. \ref{sect:candidate_sources}). The discussed
object \clumpi\space serves as an appropriate benchmark, because the
corresponding uncertainties are representative for a typical DM
subhalo source. The observational uncertainties are of similar
magnitude as the theoretical ones.

\subsection{Remarks and prospects for IACTs} \label{sect:remarks_IACTs}
The 28-month data of \textit{Fermi}-LAT contains no additional photon
detected around the nominal position. This lowers the probability of
steadiness to $\sim 25\%$ and may indicate variability, which supports
a BL Lac scenario. Vice versa, such a behaviour would also be
anticipated by a selection bias: If the true flux is lower than the
value found in the discovery data set, the discovery condition would
only allow for the detection of sources where the photon number has
been fluctuating upwards. Poisson fluctuations of this faint signal
could have accounted for a detection with the LAT even if the true
flux had remained below the detection sensitivity.

\begin{table}[t]
\caption{\label{tab:iacts}Fluxes above the energy thresholds of MAGIC
  and H.E.S.S., predicted by a DM scenario of \clumpi.}
 \begin{center}
 \begin{tabular}{lccc}
\hline \hline
\multicolumn{4}{c}{Flux prediction for MAGIC/H.E.S.S. [\%Crab]} \\
$m_\chi$ & 150\,GeV & 500\,GeV & 1\,TeV \\
\hline
$b\overline{b}$ &  & $0.3/10^{-3}$ & 0.6/0.05 \\
$W^+W^-$ &  & 0.5/0.01 & 0.8/0.2 \\
$\tau^+\tau^-$ & 0.7/-- & 3.1/1.1 & \\
\hline
 \end{tabular}
\tablefoot{The fluxes are listed in percentages of the Crab Nebula's
  flux, $\phi_\mathrm{Crab}(> 50\,\mathrm{GeV}) \approx 1.6\times
  10^{-9}\,\mathrm{cm}^{-2}\,\mathrm{s}^{-1}$
  \citep{2008ApJ...674.1037A} and
  $\phi_\mathrm{Crab}(>300\,\mathrm{GeV}) \approx 1.5\times
  10^{-10}\,\mathrm{cm}^{-2}\,\mathrm{s}^{-1}$
  \citep{2006A&A...457..899A}, respectively. Effective cross sections
  required by the individual DM scenarios are discussed in
  Sect. \ref{sect:subhalo_scenario}, raised by a factor of about 2.3
  for $m_\chi = 1\,\mathrm{TeV}$.}
 \end{center} 
\end{table}

It is instructive to note that with regard to a definite
identification of a counterpart (or ruling out a candidate) from
observations in other wavelength regimes the limiting factor is the
accuracy of the \textit{Fermi}-LAT source position
($\mathcal{O}(5^\prime)$, cf., Table \ref{tab:clump_candidates}) and
PSF. With just six detected photons, probably including one background
photon, the source is close to the confusion limit. This situation can
only be resolved by future instruments with much larger effective
areas, such as the proposed CTA, which will probe deep into the
expected population of subhaloes. The much larger number of photons
would help to infer significantly improved source
positions. Furthermore, for detecting a spectral cut-off and in case
of heavy WIMPs ($m_\chi > 1\,\mathrm{TeV}$), observations in the VHE
range with IACTs are favoured. For the particular DM scenarios
proposed for \clumpi, fluxes anticipated in the energy ranges
accessible for MAGIC and H.E.S.S., $\phi(> 50\,\mathrm{GeV})$ and
$\phi(> 300\,\mathrm{GeV})$, respectively, are listed in Table
\ref{tab:iacts} (given by $\phi(>E)\propto
N_\gamma(>E)/N_\gamma(10\!-\!100\,\mathrm{GeV})$, see
Eq. \ref{eq:boost}). Additionally, flux estimates for WIMPs with
$m_\chi=1\,\mathrm{TeV}$ were derived. Note that the required
effective cross sections (see Sect. \ref{sect:subhalo_scenario})
increase by a factor of 2.3, because \mbox{$\sigmav_\mathrm{eff} \propto
m_\chi^2\,N_\gamma(10\!-\!100\,\mathrm{GeV})^{-1}$}. Also note that fluxes
expected for VERITAS are comparable to those for H.E.S.S.

The low energy threshold of MAGIC leads to comparatively high
integrated VHE fluxes for $m_\chi < 1\,\mathrm{TeV}$. The flux
prediction for MAGIC is of $\mathcal{O}(1\%)$ of the Crab Nebula's for
the favoured $\tau^+\tau^-, m_\chi = 150\,\mathrm{GeV}$ and $W^+W^-,
m_\chi = 1\,\mathrm{TeV}$ model. With MAGIC, 50\,hours of observation
are necessary to detect this source with more than 5$\sigma$. For
comparison, predicted fluxes for H.E.S.S. are not higher than
0.2\%\,Crab for these models, which requires a few hundred hours of
observation \citep{2006A&A...457..899A}. We remark that advanced
analysis methods improve the sensitivity of H.E.S.S. by a factor of 2
\citep{2009APh....32..231D}. In the near future, an additional
telescope (H.E.S.S.-II) will lower the energy threshold of H.E.S.S. to
about 25-50\,GeV. For the corresponding flux level of 1\%\,Crab, the
required observation time for H.E.S.S.-II and MAGIC will be
similar. Furthermore, the planned CTA observatory will be able to
detect such a source in about 50\,hours \citep{2010arXiv1008.3703C}.

\section{Summary and conclusions} \label{sect:conclusion}
Hierarchical structure formation predicts \mbox{Milky Way-sized}
galaxies to host numerous DM subhaloes with masses ranging from
$10^{10}$ down to a cut-off scale of
$10^{-3}-10^{-11}\,\mathrm{M}_\odot$. Given standard WIMP scenarios,
e.g., motivated by supersymmetry, we have demonstrated that DM
subhaloes are detectable with the currently operating $\gamma$-ray
telescope \textit{Fermi}-LAT. Based upon state-of-the-art models,
detectable subhaloes would observationally appear as faint high-energy
$\gamma$-ray sources between 10 and 100\,GeV with a flux at the
sensitivity level of \textit{Fermi}-LAT (\mbox{$\sim
  10^{-10}\,\mathrm{cm}^{-2}\,\mathrm{s}^{-1}$} between 10 and
100\,GeV for one year). The observable $\gamma$-ray emission exhibits
a moderate spatial extent below \mbox{$\sim 0.5^\circ$}. Subhaloes
favoured for detection are massive ($10^5 - 10^{8}\,\mathrm{M}_\odot$)
at distances of $\mathcal{O}(\mathrm{kpc})$, while low-mass subhaloes
are not detectable. Within the intrinsic halo-to-halo scatter, only a
moderate enhancement of the self-annihilation cross section preferred
by standard cosmology, $\sigmav_0 =
3\times10^{-26}\,\mathrm{cm}^3\,\mathrm{s}^{-1}$, between 3 and 12 is
necessary (dependent on the WIMP model), which is consistent with
current observational constraints. Increasing sensitivity for a
data-taking period of five years will allow us to resolve subhaloes
requiring a cross section enhanced by a factor between 1.3 and
5. Additional sub-substructure within a subhalo may lower the required
enhancement. Within statistics, \textit{one} massive subhalo could be
detectable with \textit{Fermi}-LAT in one year and might appear in the
first-year catalogue (1FGL), assuming a subhalo population predicted
by numerical $N$-body simulations. Regarding the 1FGL, the high-energy
flux ($10\!-\!100$\,GeV) of subhalo candidates should be fainter than
$\sim 4\times 10^{-10}\,\mathrm{cm}^{-2}\,\mathrm{s}^{-1}$ (for the
WIMP scenarios considered here).

Intensive searches for subhaloes in the 1FGL reveal twelve candidates,
which are unassociated, non-variable, high-latitude sources detected
above 10\,GeV. The physical origin of the most promising object
selected by lacking association, faintness, and spectral index,
\clumpi, was investigated by analysing the 24-month data set of
\textit{Fermi}-LAT. With dedicated \textit{Swift}-XRT observations
(10.1\,ks), seven X-ray sources were discovered around
\clumpi. Located within the positional uncertainty of the $\gamma$-ray
source, a radio source positionally coincident with a newly discovered
X-ray source hints at a conventional HBL origin of \clumpi. However,
owing to a large positional uncertainty and the lacking detection of
variability, the possibility of a dark nature remains. The measured
high-energy flux and spatial extent of the source is compatible with a
DM subhalo between $10^6$ and $10^8\,\mathrm{M}_\odot$ in a distance
of about 2\,kpc, driven by a 500\,(150)\,GeV WIMP self-annihilating to
$b\overline{b}$ ($\tau^+\tau^-$). In this case, the required
enhancement of $\sigmav_0$ is 7 (3) within the intrinsic scatter of
the subhalo model, given a subhalo of $10^6\,\mathrm{M}_\odot$.

Establishing the -- probably more likely -- HBL scenario of
\clumpi\space requires a significant detection of $\gamma$-ray
variability and a confirmation of the radio as well as X-ray
counterparts. Vice versa, a steady $\gamma$-ray flux with a spectral
shape predicted by self-annihilating WIMPs would hint at a DM nature
of the object. This validates the necessity of additional intense and
long multi-wavelength observations. In particular, IACTs offer a
unique capability to reduce the positional uncertainty of faint LAT
sources and to detect a spectral cut-off in the VHE range. A detection
of the subhalo candidate \clumpi\space may be possible with telescope
systems like H.E.S.S.-II, MAGIC, and CTA within about 50\,hours of
observation.

Our results encourage the search for more subhalo candidates in
current and upcoming \mbox{(very-)high-energy} data releases. However,
even in optimistic scenarios the expected number of LAT-detectable
subhaloes is small. Furthermore, a longer exposure time -- while
certainly helpful with regard to the single candidate discussed in
this work -- will not neccessarily remedy the general problem of the
$\gamma$-ray photon count that limits the positional accuracy and
therefore the chance of identifying counterparts. Given $m_\chi <
1\,\mathrm{TeV}$, acquiring a sufficiently large number of detections
which may solve the subhalo problem requires higher sensitivity in the
high-energy range.

\begin{acknowledgements}
We kindly acknowledge helpful discussions with our colleagues
Katharina Borm, Torsten Bringmann, Wilfried Buchm\"uller, Frederike
J\"ager, Andrei Lobanov, and Martin Raue. We kindly acknowledge the
\textit{Swift} PI Neil Gehrels and his team for the prompt response to
our ToO request and the corresponding observations. The help of the
Fermi HelpDesk is kindly acknowledged. We kindly thank the anonymous
referee for useful comments. This work was supported through the
collaborative research center (SFB) 676 ``Particles, Strings, and the
Early Universe'' at the University of Hamburg.
This publication makes use of data products from the Two Micron All
Sky Survey, which is a joint project of the University of
Massachusetts and the Infrared Processing and Analysis
Center/California Institute of Technology, funded by the National
Aeronautics and Space Administration and the National Science
Foundation.
This publication makes use of data from the Sloan Digital Sky Survey
(SDSS). Funding for the SDSS and SDSS-II has been provided by the
Alfred P. Sloan Foundation, the Participating Institutions, the
National Science Foundation, the U.S. Department of Energy, the
National Aeronautics and Space Administration, the Japanese
Monbukagakusho, the Max Planck Society, and the Higher Education
Funding Council for England. The SDSS Web Site is
http://www.sdss.org/.
The SDSS is managed by the Astrophysical Research Consortium for the
Participating Institutions. The Participating Institutions are the
American Museum of Natural History, Astrophysical Institute Potsdam,
University of Basel, University of Cambridge, Case Western Reserve
University, University of Chicago, Drexel University, Fermilab, the
Institute for Advanced Study, the Japan Participation Group, Johns
Hopkins University, the Joint Institute for Nuclear Astrophysics, the
Kavli Institute for Particle Astrophysics and Cosmology, the Korean
Scientist Group, the Chinese Academy of Sciences (LAMOST), Los Alamos
National Laboratory, the Max-Planck-Institute for Astronomy (MPIA),
the Max-Planck-Institute for Astrophysics (MPA), New Mexico State
University, Ohio State University, University of Pittsburgh,
University of Portsmouth, Princeton University, the United States
Naval Observatory, and the University of Washington.
This research has made use of the NASA/IPAC Extragalactic Database
(NED) which is operated by the Jet Propulsion Laboratory, California
Institute of Technology, under contract with the National Aeronautics
and Space Administration.
\end{acknowledgements}

\begin{appendix}
\section{Concentration of Aquarius subhaloes} \label{app:cvir_Aq}
The Aquarius simulation provides results on the profile parameters of
resolved subhaloes, taking tidal interaction into account
\citep{2008MNRAS.391.1685S}. These results are used to derive the
distance-averaged virial concentration of subhaloes to confront it
with the toy-model predictions used here.

Following up on Eq.~\ref{eq:sub_profile}, the \textit{tidal}
concentration $c_\mathrm{t} \equiv R_\mathrm{t}/r_\mathrm{s}$ is
introduced \citep[cf.,][]{2009PhRvD..80b3520A}, where $R_\mathrm{t}$
denotes the tidal and therefore physical radius of a subhalo. For an
NFW-type mass density profile, $c_\mathrm{t} = \exp[W(-e^{-a})+a] -1,
a \equiv 1 + M_\mathrm{t}/(4\pi \rho_\mathrm{s} r_\mathrm{s}^3)$,
where $W(x)$ denotes Lambert's W-function and $M_\mathrm{t}$ the tidal
subhalo mass. In numerical simulations, the directly ''observable''
quantities of (sub)haloes are related to the dynamics of the halo
system, including the maximum velocity $V_\mathrm{max}$ and the
distance $r_\mathrm{max}$ where $V_\mathrm{max}$ is reached. To
recover the canonical parameters $r_s$ and $\rho_s$ related to the
density profile, we use approximate relations $2 [V_\mathrm{max}/(H_0
  r_\mathrm{max})]^2 \simeq 5.80\times
10^4\,[M_\mathrm{t}/(10^8\,\mathrm{M}_\odot)]^{-0.18}$ and
$M_\mathrm{t} \simeq 3.37 \times 10^7\,
[V_\mathrm{max}/(10\,\mathrm{km}\,\mathrm{s}^{-1})]^{3.49}\,\mathrm{M}_\odot$,
fitting the results of the simulation
\citep{2008MNRAS.391.1685S,2009PhRvD..80b3520A}. Given analytical
relations between $(r_\mathrm{max},V_\mathrm{max})$ and
$(r_\mathrm{s},\rho_\mathrm{s})$ for the NFW profile \citep[e.g.,
  Eq. 8 in][]{2008ApJ...686..262K}, this yields
$r^\mathrm{Aq}_\mathrm{s}(M_\mathrm{t}) \simeq 0.094\,[
  M_\mathrm{t}/(10^6\,\mathrm{M}_\odot) ]^{0.38}\,\mathrm{kpc}$ and
$\rho^\mathrm{Aq}_\mathrm{s}(M_\mathrm{t}) \simeq 9.6 \times 10^5
\rho_\mathrm{crit}\,[ M_\mathrm{t}/(10^6\,\mathrm{M}_\odot)
]^{-0.18}$. Therefore, the tidal concentration
$c_\mathrm{t}^\mathrm{Aq}$ is determined via $a^\mathrm{Aq} \simeq 1 +
0.66\,[M_\mathrm{t}/(10^6\,\mathrm{M}_\odot)]^{0.04}$, which is valid
for masses above the resolution limit of the simulation, $M_\mathrm{t}
\gtrsim 3.2\times 10^4\,\mathrm{M}_\odot$.

The virial concentration of Aquarius subhaloes is given by
\mbox{$c_\mathrm{vir}^\mathrm{Aq}(M_\mathrm{vir}) = [3
    M_\mathrm{vir}/(4\pi \Delta_\mathrm{c}
    \rho_\mathrm{crit})]^{1/3}/r^\mathrm{Aq}_\mathrm{s}(M_\mathrm{vir})$},
where the characteristic radius as function of the virial subhalo mass
is obtained from an empirical relation mapping $M_\mathrm{vir}$ to
$M_\mathrm{t}$. Based on $c_\mathrm{t}^\mathrm{Aq}$ and assuming the
FHM virial concentration-to-mass relation (Eq. \ref{eq:cvir_lavalle}),
the relative tidal mass is \mbox{$M_\mathrm{t}/M_\mathrm{vir} \approx
  f[c^\mathrm{Aq}_\mathrm{t}(M_\mathrm{t})]/f[c^\mathrm{FHM}_\mathrm{vir}(M_\mathrm{vir})]$},
since $M = 4\pi \rho_\mathrm{s} r_\mathrm{s}^3 f(c)$. Hereby, we have
assumed that tidal effects on inner subhalo parts are negligible:
$\rho_\mathrm{s}(M_\mathrm{vir}) r_\mathrm{s}(M_\mathrm{vir})^3
\approx \rho_\mathrm{s}(M_\mathrm{t})
r_\mathrm{s}(M_\mathrm{t})^3$. With $f(c^\mathrm{Aq}_\mathrm{t}) =
a^\mathrm{Aq}-1$, the distance-averaged
\mbox{$M_\mathrm{t}$-$M_\mathrm{vir}$} relation is
\begin{equation} \label{eq:app_Mt_Mv}
 M_\mathrm{t}(M_\mathrm{vir}) \simeq \left(
 \frac{712.6\,\mathrm{kpc}^{-3}}{4\pi \rho_\mathrm{crit}}
 \right)^{1.04} \left(
 \frac{M_\mathrm{vir}}{f(c^\mathrm{FHM}_\mathrm{vir})} \right)^{1.04}
 \mathrm{M}_\odot.
\end{equation}
For massive subhaloes \mbox{($\gtrsim 10^4\,\mathrm{M}_\odot$)},
$f(c^\mathrm{FHM}_\mathrm{vir})^{-1.04}$ is well fit by a power law,
$f[c^\mathrm{FHM}_\mathrm{vir}(M_\mathrm{vir})]^{-1.04} \approx
0.34\,[M_\mathrm{vir}/(10^6\,\mathrm{M}_\odot)]^{0.02}$, yielding
$M_\mathrm{t}/M_\mathrm{vir} \approx 0.23
[M_\mathrm{vir}/(10^6\,\mathrm{M}_\odot)]^{0.06}$ for $M_\mathrm{t}
\gtrsim 3.2\times 10^4\,\mathrm{M}_\odot$. This reveals
$r^\mathrm{Aq}_\mathrm{s}(M_\mathrm{vir}) \simeq 0.054\,[
  M_\mathrm{vir}/(10^6\,\mathrm{M}_\odot) ]^{0.40}\,\mathrm{kpc}$ and,
therefore, the distance-averaged virial concentration of subhaloes
\begin{equation}
c^\mathrm{Aq}_\mathrm{vir}(M_\mathrm{vir}) \simeq
46.8\,\left(\frac{M_\mathrm{vir}}{10^6\,\mathrm{M}_\odot}\right)^{-0.07}
\end{equation}
for $M_\mathrm{vir} \in [1.5\times 10^5;\sim 10^{10}]\,\mathrm{M}_\odot$.
\end{appendix}

\begin{appendix}
\section{Moderately extended \textit{Fermi} sources} \label{app:mc}
For $\gamma$-ray catalogues such as 1FGL, instrument data have been
analysed assuming sources to be point-like. Given that detectable
subhaloes would appear as moderately extended according to the PSF of
\textit{Fermi}-LAT (see Sect. \ref{sect:fiducial_candidate:obsprop},
$\sigma_\mathrm{PSF} \approx 0.15^\circ$ for $E = 10$\,GeV), we
investigated the effect of the 1FGL point-source-analysis framework on
extended sources.

To study the high-energy flux $\phi_\mathrm{p}(10\!-\!100\,\mathrm{GeV})$
reconstructed by the point-source analysis for a given intrinsic
(subhalo) extent $\theta_\mathrm{s}$, a Monte-Carlo (MC) simulation
dedicated to the particular source \clumpi\space was used. Based
on the 11-month data set (see Sect. \ref{sect:fermi_data} for
details), the celestial coordinates of each of the five source photons
between 10 and 100\,GeV were re-simulated. The intensity profile
was assumed to follow the line-of-sight integral over the
(squared) NFW profile of a subhalo for the given
$\theta_\mathrm{s}$ (peaking at the nominal source position). Other
observational photon parameters, such as energy, inclination,
detection time, conversion type, and event class (see Table
\ref{tab:photons}), were kept fixed. Subsequently, detectional
influences were accounted for by smoothing with the PSF. For each
$\theta_\mathrm{s}$, 500 iterations were analysed with the
framework described in Sect. \ref{sect:fermi_data} (\textit{gtfindsrc}
and \textit{gtlike}) according to flux and significance ($S \approx
\sqrt{\mathrm{TS}}$, where TS denotes the test statistic of the
analysis). All other sources within the ROI were kept fixed. The
study is restricted to the signal-dominated regime chosen to be
$\theta_\mathrm{s} \lesssim 1^\circ$ given the low background
$N_\mathrm{bg}$. Since $\theta_{68} \approx 0.46^\circ$, this
corresponds to $\sim 3\, \sigma_\mathrm{PSF}$. Justified by the low
background, all photons were treated as signal events.

\begin{figure}[t]
  \resizebox{\hsize}{!}{\includegraphics{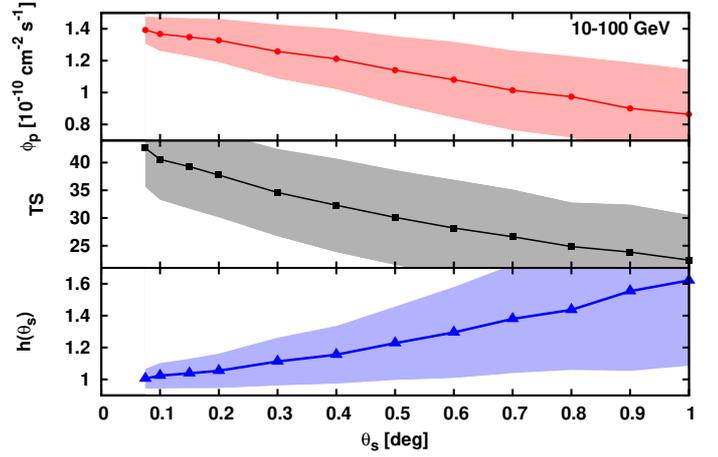}}
  \caption{Average (solid lines) and rms (shaded areas) of
    \mbox{$\phi_\mathrm{p}(10\!-\!100\,\mathrm{GeV})$} (upper panel), TS (middle
    panel), and the scaling $h(\theta_\mathrm{s})$ (lower panel) as
    function of the intrinsic angular extent $\theta_\mathrm{s}$. For
    each $\theta_\mathrm{s}$, a sample of 500 simulations of the
    \clumpi\space photon distribution between 10 and 100 GeV assuming
    a DM subhalo intensity profile was analysed with the 1FGL
    point-source-analysis framework.}
  \label{fig:sumMC}
\end{figure}

The $\theta_\mathrm{s}$ dependence of the sample-averaged
reconstructed flux $\phi_\mathrm{p}(10\!-\!100\,\mathrm{GeV})$ and
corresponding test statistic TS is shown in the two upper panels of
Fig. \ref{fig:sumMC}. For large $\theta_\mathrm{s}$, the probability
of photons to be located far away from their central position
increases. Therefore, both $\phi_\mathrm{p}$ and TS decrease because
of a minor contribution of outer photons to the point-source region
(defined by the PSF). For $\theta_\mathrm{s} \approx 1^\circ$, the
average significance drops below the detection criterion ($\mathrm{TS}
\geq 25$). Note that $\mathrm{TS} \geq 25$ still holds for about 35\%
of the simulated samples.

In terms of Eq. \ref{eq:boost}, appropriate investigation of
candidates provided by point-source catalogues is therefore admitted
by a scaling $h(\theta_\mathrm{s})$, which allows us to map the
catalogued flux $\phi_\mathrm{p}$ to the true flux $\phi$ of the
entire source. The angular dependence of $h$ is shown in the lower
panel of Fig. \ref{fig:sumMC}. Given $\phi =
h(\theta_\mathrm{s})\,\phi_\mathrm{p}$, the factor was derived by
defining $h(0^\circ) = 1$. Conservatively, the complete MC sample was 
used to derive $h(\theta_\mathrm{s})$, including realisations
with $\mathrm{TS}<25$\,\footnote{Given the selection bias of the 1FGL
  catalogue, $\mathrm{TS} \geq 25$, a more stringent deduction of
  $h(\theta_\mathrm{s})$ should include realisations with $\mathrm{TS}
  \geq 25$ only. This lowers the effective scaling factor
  $h(\theta_\mathrm{s})$.}. As expected in the signal-dominated
regime, the increase of $h$ with increasing $\theta_\mathrm{s}$ is
comparatively slight, while it is fairly linear in the
background-dominated regime. Note again that this result holds for
sources similar to \clumpi\space at high galactic latitudes only,
while in general $h = h(l,b,\theta_\mathrm{s})$.

Vice versa, Fig. \ref{fig:sumMC} states a reasonable (but
conservative) value of the sensitivity of \textit{Fermi}-LAT for hard
sources of similar type: $\phi_\mathrm{p}(10\!-\!100\,\mathrm{GeV})
\approx 10^{-10}\,\mathrm{cm}^{-2}\,\mathrm{s}^{-1}$. Note that this
value is similar to the point-source sensitivity stated in
\citet{2009ApJ...697.1071A}.
\end{appendix}

\begin{appendix}
\section{Subhalo-induced diffuse flux} \label{app:sub_diff}
In the following, the diffuse flux of the subhalo population is
derived using a prescription by \citet{2009PhRvD..80b3520A}, which is
extended to include the probability distribution of the concentration
parameter $c$ (see Eq.~\ref{eq:scatter}). Numerical $N$-body
simulations have demonstrated that the differential subhalo number
density $\mathrm{d}n_\mathrm{sh}=\mathcal{N}(r,M)\,\mathrm{d}M$
follows a power-law in subhalo mass $M$. Following standard
assumptions, the number density $\mathcal{N}(r,M)$ factorises such
that $\mathcal{N}(r,M) \propto n_\mathrm{sh}(r)\cdot M^{-\alpha}$,
where $\alpha = 1.9$ and $r$ is the distance to the host's centre. In
simulations, the spatial density distribution $n_\mathrm{sh}(r)$ is
consistently found to be ``anti-biased'' and, e.g., $n_\mathrm{sh}(r)
\propto \rho_\mathrm{Ein}(r)$ \citep{2008MNRAS.391.1685S}, where
$\rho_\mathrm{Ein}(r)$ denotes the Einasto profile \citep{1965Einasto}
\begin{equation}
 \rho_\mathrm{Ein}(r) \propto \exp\left\{-\frac{2}{\alpha_\mathrm{E}}
 \left[ \left( \frac{r}{r_{-2}} \right)^{\alpha_\mathrm{E}} - 1
   \right] \right\}.
\end{equation}
For a Milky Way-sized halo, the best-fit parameters for the subhaloes'
spatial distribution $\rho_\mathrm{Ein}(r)$ have been found to be
$\alpha_\mathrm{E} = 0.68$ and $r_{-2} =
0.81\,c^\mathrm{MW}_{200}\,r_\mathrm{s}^\mathrm{MW}$
\citep{2008MNRAS.391.1685S}, where $c^\mathrm{MW}_{200} \approx 15$
\citep{2010JCAP...08..004C}. Using $\mathcal{N}(r,M)$ normalised to
represent a probability density function in $M$, the differential
density is
\begin{equation} \label{eq:mass_spatial_distr}
 \frac{\mathrm{d}n_\mathrm{sh}(r,M)}{\mathrm{d}M}=n_\mathrm{sh}(r)\frac{\alpha-1}{M_\mathrm{min}}\left(
 \frac{M}{M_\mathrm{min}} \right)^{-\alpha},
\end{equation}
where $M_\mathrm{min} \ll M_\mathrm{max}$ are the minimum and maximum
mass of Galactic subhaloes, respectively. The normalisation of the
subhalo number density $n_\mathrm{sh}(r)$ is chosen such that the
fraction of the host's mass distributed in subhaloes $f_\mathrm{sh}
\equiv M_\mathrm{sh}/M^\mathrm{MW}_\mathrm{vir} = 15\%$ for the
cut-off scale $M_\mathrm{min} = 10^{-6}\,\mathrm{M}_\odot$, where
$M^\mathrm{MW}_\mathrm{vir} = (1.49 \pm 0.17)\times
10^{12}\,\mathrm{M}_\odot$ \citep{2010JCAP...08..004C}. The chosen
value of $f_\mathrm{sh}$ is consistent with recent estimates
$f_\mathrm{sh} = 10\!-\!50\%$
\citep{2005Natur.433..389D,2009arXiv0906.4340D,2008Natur.454..735D,2008Natur.456...73S}. The
total mass contained in subhaloes is given by
\begin{equation}
 f_\mathrm{sh} M^\mathrm{MW}_\mathrm{vir} = 4\pi
 \int_0^{R^\mathrm{MW}_\mathrm{vir}} \mathrm{d}r\,r^2
 \int_{M_\mathrm{min}}^{M_\mathrm{max}}
 \mathrm{d}M\,M\frac{\mathrm{d}n_\mathrm{sh}(r,M)}{\mathrm{d}M}.
\end{equation}
Solving for an Einasto-type profile and $\alpha \neq 2$ yields
\begin{eqnarray}
 n_\mathrm{sh}(r) & = & \frac{f_\mathrm{sh}
   M^\mathrm{MW}_\mathrm{vir}}{2\pi r^3_{-2} M_\mathrm{min}} \left(
 \frac{2}{\alpha_\mathrm{E}} \right)^{3/\alpha_\mathrm{E}-1} \Gamma
 \left[ \frac{3}{\alpha_\mathrm{E}}, \frac{2}{\alpha_\mathrm{E}}
   \left( \frac{R^\mathrm{MW}_\mathrm{vir}}{r_{-2}}
   \right)^{\alpha_\mathrm{E}}\right]^{-1} \nonumber \\ 
  & \times &
 \frac{2-\alpha}{(\alpha-1) (\Lambda^{2-\alpha}-1)} \exp \left[
   -\frac{2}{\alpha_\mathrm{E}} \left( \frac{r}{r_{-2}}
   \right)^{\alpha_\mathrm{E}} \right],
\end{eqnarray}
where $\Gamma(a,x)$ is the lower incomplete gamma function and
$\Lambda = M_\mathrm{max}/M_\mathrm{min}$.

The minimum mass $M_\mathrm{min}$ of subhaloes is governed by the
details of kinetic decoupling of WIMPs in the early Universe
\citep{2003PhRvD..68j3003B,2006PhRvD..73f3504B,2005JCAP...08..003G,2009NJPh...11j5027B}. Depending
upon the mass and composition of, e.g., the neutralino, a wide range
of minimal subhalo masses has been considered in the literature,
namely $M_\mathrm{min} \in
[10^{-11};10^{-3}]\,\mathrm{M}_\odot$. Here, two benchmark cases are
considered for $M_\mathrm{min}$, i.e., $10^{-10}$ and
$10^{-6}\,\mathrm{M}_\odot$, bracketing the 500\,GeV neutralino
scenario discussed by \citet{2009NJPh...11j5027B}. The upper mass
limit was fixed to $M_\mathrm{max} =
10^{-2}\,M^\mathrm{MW}_\mathrm{vir} \approx
10^{10}\,\mathrm{M}_\odot$. Results do not depend on the exact value
chosen for $M_\mathrm{max}$.

Using $\mathcal{L}(M,D)$ (Sect. \ref{sect:density} and
\ref{sect:luminosity}), the average specific intensity from a subhalo
population with extended, isotropic emissivity profiles is given
towards a galactic direction $\hat{\mathbi{n}}$ by
\begin{equation} \label{eq:diffint}
 \langle I_\nu(\hat{\mathbi{n}}) \rangle = \hskip -0.2cm
 \int\limits_{M_\mathrm{min}}^{M_\mathrm{max}} \hskip -0.15cm
 \mathrm{d}M \hskip -0.3cm
 \int\limits_{s_*(\mathcal{L}(M,\tilde{s}))}^{s_\mathrm{max}(\hat{\mathbi{n}})}
\hskip -0.4cm \mathrm{d}s \frac{\mathrm{d}n_\mathrm{sh}(r(s,\hat{\mathbi{n}}),M)}{\mathrm{d}M} \hskip -0.15cm
\int \hskip -0.15cm \mathrm{d}c\, P(c,\overline{c})\, \frac{\mathcal{L}_\nu(M,c)}{4 \pi},
\end{equation}
assuming that the spatial extent of each subhalo is much smaller than
the scale on which the subhalo distribution changes significantly. The
total photon rate $\mathcal{L}_\nu$ is given by
Eq. \ref{eq:luminosity_vir} with the substitution $N_\gamma
\rightarrow E\, \mathrm{d}N_\gamma/\mathrm{d}E$. Furthermore,
$\mathcal{L}_\nu(M)$ is required to be one-to-one. The galactocentric
radius corresponding to the position $s\,\hat{\mathbi{n}}$ is
$r(s,\psi) = ( R_0^2 + s^2 - 2 R_0 s \cos \psi )^{1/2}$, where $\psi$
denotes the angle between $\hat{\mathbi{n}}$ and $\hat{\mathbi{R}}_0$
($\cos(\psi)=\langle \hat{\mathbi{n}},\hat{\mathbi{R}}_0
\rangle$). Subhaloes bright enough to be detected as individual
sources are not considered to contribute to the diffuse
emission. Therefore, the lower limit of the line-of-sight integral is
set by the detection criterion $\mathcal{L} \geq 4\pi s_*^2
\phi_\mathrm{sens}$, where $\phi_\mathrm{sens}$ denotes the flux
sensitivity for a detection in one year with \textit{Fermi}-LAT, see
Sect. \ref{sect:Fermi_clumps}. Since $R_0 \ll R_\mathrm{vir}$, the
upper bound of the $s$-integral $s_\mathrm{max}(\hat{\mathbi{n}})
\approx R^\mathrm{MW}_\mathrm{vir}$. The SHM photon rate is a function
of both $M$ and $s$, and therefore $s_*$ also depends slightly on
$s$. Conservatively, $s_*(M,s) = s_*(M,\tilde{s})$, $\tilde{s} =
R^\mathrm{MW}_\mathrm{vir}$, revealing a lower bound on $s_*$.
\end{appendix}

\bibliographystyle{aa}
\bibliography{aa17655-11}

\end{document}